\documentclass[twocolumn,showpacs,preprintnumbers,amsmath,amssymb]{revtex4}

\usepackage{graphicx}% Include figure files
\usepackage{dcolumn}% Align table columns on decimal point
\usepackage{bm}% bold math
\usepackage{amsmath}
\usepackage{url}
\usepackage{color}

\begin{document}

\preprint{LA-UR-13-20684}

\title{Halo modification of a supernova neutronization neutrino burst}% Working title
\author{John F. Cherry$^{1,2,3,4}$}
\author{J. Carlson$^{1,4}$}
\author{Alexander Friedland$^{1,4}$}
\author{George M. Fuller$^{3,4}$}
\author{Alexey Vlasenko$^{3,4}$}

\affiliation{$^{1}$Theoretical Division, Los Alamos National Laboratory, Los Alamos, New Mexico 87545, USA}
\affiliation{$^{2}$Department of Physics and Astronomy, University of New Mexico, Albuquerque, New Mexico 87131, USA}
\affiliation{$^{3}$Department of Physics, University of California, San Diego, La Jolla, California 92093, USA}
\affiliation{$^{4}$Neutrino Engineering Institute, New Mexico Consortium, Los Alamos, New Mexico 87545, USA}

\date{February 4, 2013}

\begin{abstract}
We give the first self-consistent calculation of the effect of the scattered neutrino halo on flavor evolution in supernovae.  Our example case is an O-Ne-Mg core collapse supernova neutronization neutrino burst.  We find that the addition of the halo neutrinos produces qualitative and quantitative changes in the final flavor states of neutrinos.  We also find that the halo neutrinos produce a novel distortion of the neutrino flavor swap.   Our results provide strong motivation for tackling the full multidimensional and composition-dependent aspects of this
problem in the future.
\end{abstract}

\pacs{14.60.Pq, 97.60.Bw}

\maketitle

\section{Introduction}
In this paper we give the first self-consistent treatment of supernova neutrino flavor evolution that includes the \lq\lq halo\rq\rq ~\cite{Cherry:2012uq} of neutrinos generated by direction-changing scattering in the supernova envelope.  Neutrinos in the core collapse supernova environment are emitted primarily from the protoneutron star left by the collapse, which is considerably smaller than the envelope of the collapsing star.  As neutrinos propagate outward, their flavor evolution is determined, in part, by their coherent forward-scattering off of other neutrinos, almost all of which are emitted from the protoneutron star.  However, the neutrino-neutrino coherent forward scattering contribution to the potential that governs flavor transformation is sensitive to the intersection angle of interacting neutrinos, such that the resultant potential is $\propto \left( 1-\cos\theta\right)$, where $\theta$ is the angle between the incident neutrino trajectories.  This proportionality has two direct consequence for the interactions of neutrinos in the supernova explosion.  Neutrinos emerging from the protoneutron star will experience a significantly suppressed self-interaction once they have propagated more than a few protoneutron star radii.  Neutrinos that have scattered at wide angles in the outer envelope of the explosion (halo neutrinos), while considerably less numerous than neutrinos emerging form the core, experience no such suppression of the neutrino-neutrino forward scattering potential.  This lack of geometric suppression leads to the result that during the first $\sim 1\,\rm s$ of a core collapse supernova explosion, halo neutrinos can be the dominant source of neutrino-neutrino forward scattering potential in regions where active neutrino flavor transformation may take place~\cite{Cherry:2012uq}.

Taken at face value, the halo changes the nature of flavor evolution calculations, converting them from initial value problems into boundary value problems.  The reason for this is that direction-changing scattering, in principle, can cause neutrino flavor information to propagate inward from a relatively large radius.  Other studies of the halo have concentrated on stability of the neutrino flavor field in the accretion phase/shock reheting epoch of the supernova~\cite{Sarikas:2012qy}, where composition, hydrodynamics-generated matter inhomogeneity, as well as inwardly propagating neutrinos currently preclude self-consistent calculations.  In contrast, the very compact and centrally concentrated nature of the matter density distribution in O-Ne-Mg core collapse supernovae makes the halo in these cases amenable to a self-consistent initial value treatment.  Here we exploit this felicitous feature of O-Ne-Mg core collapse and thereby take a step toward a more comprehensive treatment of neutrino transport.

In fact, a longstanding and unresolved question in the physics of core collapse supernovae is the simultaneous and self-consistent solution of neutrino transport and flavor transformation.  This subject has historically been approached by splitting the treatment of neutrinos in supernovae into two distinct limits: the Boltzmann transport limit, which contains all of the physical processes by which neutrinos are created, absorbed, and scattered in new directions; and the coherent forward-scattering limit, which governs the evolution of neutrino flavor states that are freely streaming.  However, the existence of the neutrino halo calls into question this separation that is at the heart of the current neutrino transport/flavor evolution paradigm. 

The halo shows that these two limits of neutrino transport confabulate in the general case.  The neutrino halo effect itself is driven by the collusion of Boltzmann neutrino transport together with the coherent forward scattering of neutrinos.  The geometric structure of coherent neutrino-neutrino forward scattering strongly suppresses the interaction of neutrinos that are propagating on nearly co-linear trajectories, while enhancing the interaction of neutrinos that cross paths at large intersection angles.  Neutrino-nucleon/nucleus neutral current, isoenergetic interactions produce a population of neutrinos that may have scattered at wide angles, i.e. the halo neutrinos.  This latter effect is dependent on the density and composition of matter in the envelope because of the coherent enhancement of neutrino scattering on heavy nuclei~\cite{Freedman:1974yq,Tubbs:1975ve} and, consequently, couples the halo neutrino population to the hydrodynamic and nuclear evolution of the supernova as a whole.  Taken together, coherent forward scattering and neutral current direction-changing scattering can give halo neutrinos disproportionate weight in determining the flavor evolution {\it of all neutrinos}~\cite{Cherry:2012uq}.

At first glance, it may seem straightforward to include the halo neutrino population in calculations of neutrino flavor transformation, but the wide angle scattering that characterizes the trajectories of the halo neutrinos also makes their inclusion in such calculations difficult.  Relativistic Boltzmann transport of neutrinos has already been implemented in two and three spatial dimensions in simulations of core collapse supernovae~\cite{Bruenn:2010qy,Muller:2010kx,Brandt:2011lr,Muller:2012qy,Bruenn:2013fk}.  This multi-dimensional transport treatment benefits from the fact that the relevant length scale for this kind of neutrino scattering is similar to the length scale relevant to resolving the hydrodynamic evolution of the envelope.  Unlike Boltzmann neutrino transport, the neutrino coherent forward scattering limit demands that the complex phases of neutrino flavor state  high frequency oscillations be followed on length scales as small as $\sim \left(\rm a\ few\right)\, \rm cm$.  The very short length scales associated with self-consistently following neutrino flavor evolution have stymied efforts to expand simulations to multiple spatial dimensions.  Current state of the art calculations in this field are predicated on the assumption that the only neutrino states that need to be followed are outwardly-directed.  The discovery of the neutrino halo has shown this assumption to be untrustworthy during the core collapse explosion epoch, and appropriate only during the post-explosion neutrino driven wind epoch.  

This does not mean, however, that the effects of the neutrino halo on flavor transformation are outside the realm of consideration.  The low mass of the O-Ne-Mg core collapse supernova progenitors results in a prompt and spherically symmetric explosion.  This allows us to use our extant, spherically symmetric and multi-angle flavor transformation calculations for this particular case at early times.  Further, a calculation of neutrino flavor evolution can be made so long as there are few enough halo neutrinos on inwardly-directed trajectories that they do not contribute significantly to forward scattering potentials (as is the case during the neutrino driven wind epoch)~\cite{Cherry:2012uq}.  Finally, there must be some reasonable expectation that all neutrino flavor states can be specified from the start of the calculation.

The neutronization burst stage of the O-Ne-Mg core collapse supernova satisfies these criteria because of its centrally condensed matter envelope. 
\begin{figure}
\centering
\includegraphics[scale=.70]{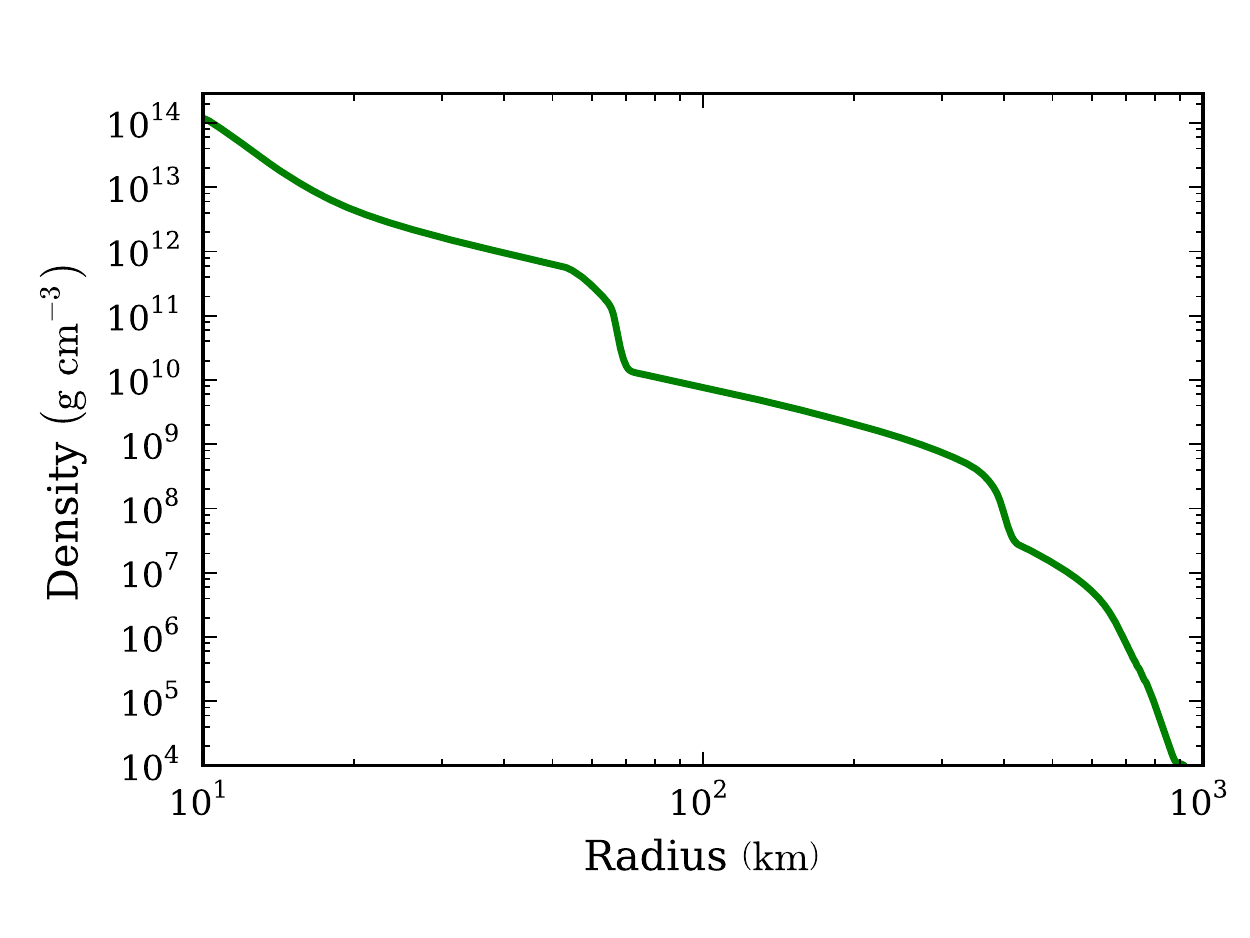}
\caption{The matter density profile taken from Reference~\cite{Fischer:2010lq}.  The profile is quite centrally concentrated, so much so that the scattered halo is only populated with an appreciable number of neutrinos inside a radius of $\sim 1000\,\rm km$.}
\label{fig:1Ddens}
\end{figure} 
Figure~\ref{fig:1Ddens} shows the density versus radius profile for the O-Ne-Mg configuration of Reference~\cite{Fischer:2010lq,Wanajo:2011lr}.  The region of the envelope that provides significant numbers of neutrinos to the halo population is located within a radius of $\sim 1000\,\rm km$~\cite{Cherry:2012uq}.  At the edge of this region the matter density drops precipitously, and ceases to scatter enough neutrinos into the halo to make a significant contribution to the forward scattering potentials.  A calculation of flavor transformation that begins outside this point may safely neglect halo neutrinos on inwardly-directed trajectories.

Previous studies have shown that flavor transformation in the O-Ne-Mg neutronization burst does not proceed until a radius of $\sim 1100\,\rm km$~\cite{Duan08,Lunardini08,Cherry:2010lr}.  This is due to the large flux of neutrinos emitted during the neutronization burst.  What this means for this particular case, where no flavor transformation has taken place inside the region which scatters neutrinos into the halo, is that the flavor states of all of the halo neutrinos can be determined from the flavor states of neutrinos emerging from the core.  This allows us to perform our spherically symmetric, multi-angle, flavor transformation calculations by starting at a radius where backwards going neutrinos are negligible.

In what follows we discuss the methodology of our calculation of in Section II, and results in Section III.  We give conclusions in Section IV.

\section{Methodology}
For the neutrino emission during the neutronization neutrino burst, we use the results of Reference~\cite{Fischer:2010lq} to set the spectral energy distribution of all flavors of neutrinos.  The neutrino emission parameters for each time slice we consider are show in table~\ref{tab:Emission}. 
\begin{table}
    \begin{tabular}{ | l | l | l |}
    \hline
     & $\displaystyle t_{\rm post\ bounce} = 7\,{\rm ms}$ & $\displaystyle t_{\rm post\ bounce} = 15\,{\rm ms}$ \\ \hline
     $\displaystyle L_{\nu_{\rm e}} $ & $\displaystyle 3.3\times10^{53}\,{\rm erg}\,{\rm s}^{-1}$ & $\displaystyle 1.3\times10^{53}\,{\rm erg}\,{\rm s}^{-1}$ \\ \hline
    $\displaystyle L_{\bar\nu_{\rm e}}  $ & $\displaystyle 2.6\times10^{51}\,{\rm erg}\,{\rm s}^{-1}$ & $\displaystyle 9.1\times10^{51}\,{\rm erg}\,{\rm s}^{-1}$ \\
    \hline
    $\displaystyle L_{\nu_{\mu / \tau},\,\bar\nu_{\mu / \tau}} $ &$\displaystyle 1.6\times10^{52}\,{\rm erg}\,{\rm s}^{-1}$ & $\displaystyle 2.6\times10^{52}\,{\rm erg}\,{\rm s}^{-1}$ \\ \hline
    $\displaystyle \langle E_{\nu_{\rm e}}\rangle$ & $\displaystyle 13.0\,\rm MeV$& $\displaystyle 11.3\,\rm MeV$ \\ \hline
   $\displaystyle \langle E_{\bar\nu_{\rm e}}\rangle$ & $\displaystyle 9.8\,\rm MeV$ & $\displaystyle 10.6\,\rm MeV$ \\ \hline
   $\displaystyle \langle E_{\nu_{\mu / \tau},\,\bar\nu_{\mu / \tau}}\rangle$ & $\displaystyle 16.7\,\rm MeV$ & $\displaystyle 15.4\,\rm MeV$ \\ \hline
    \end{tabular}
\caption{Neutrino emission parameters for the initial spectra used in our calculations. }
\label{tab:Emission}
\end{table}
We use these parameters to fit the total neutrino emission to a normalized Fermi-Dirac spectrum for each flavor of neutrino and anti-neutrino
\begin{equation}
f_{\nu}(E)=\frac{1}{F_2(\eta_{\nu})T_{\nu}^3}
\frac{E^2}{\exp(E/T_{\nu}-\eta_{\nu})+1},
\end{equation}
where we take $\eta_{\nu}=3$ and $T_{\nu}= F_2(\eta_{\nu})\langle E_{\nu}\rangle/F_3(\eta_{\nu})$.  Here
\begin{equation}
F_n(\eta)=\int_0^\infty\frac{x^n}{\exp(x-\eta)+1}dx.  
\end{equation}
For the purposes of this study we have chosen the following neutrino mixing parameters: neutrino mass squared differences $\Delta m^{2}_{\odot} = 7.6\times10^{-5}\, \rm eV^{2}$ and $\Delta m^{2}_{\rm atm} = 2.4\times10^{-3}\, \rm eV^{2}$; vacuum mixing angles $\theta_{12} = 0.59$, $\theta_{23} = \pi/4$, $\theta_{13} = 0.152$; and CP-violating phase $\delta = 0$.  

Fundamentally, the calculation that we are performing in this case is an initial value problem specified at a fixed initial radius.  To begin, we specify a unique intensity and spectral energy distribution for each flavor of neutrino and anti-neutrino on a surface of fixed radius, and simultaneously solve the non-linearly coupled equations of motion for the evolution of neutrino flavor states in the coherent forward scattering limit.  The requirement of simultaneity in the calculation of neutrino flavor states forces the entire solution to move outward, in lock-step along the radial coordinate, for all neutrinos.  By way of contrast, the neutrino halo is a manifestly multi-dimensional phenomenon, with neutrinos moving in all directions.  Including the halo in an initial value problem formalism such as ours raises a number of thorny issues that must be addressed.  

The most stringent requirement is that the calculation must be limited entirely to the region where halo neutrinos propagating inward, {\it opposite} to the direction of the calculation itself, are truly negligible.  So that we may compare the effect of the halo neutrinos to previous work, we employ the density  and composition profile found in Reference~\cite{Fischer:2010lq}, which is fit to the collapse progenitor profiles of Refs.~\cite{Nomoto84,Nomoto87} when flavor transformation calculations extend past the outer radial coordinate of the profile of Reference~\cite{Fischer:2010lq}.  To compute the neutral current neutrino scattering off of nucleons and nuclei we again use the energy dependent cross sections found by Reference~\cite{Tubbs:1975ve}.  For any scattering that takes place outside the simulation volume of Reference~\cite{Fischer:2010lq}, and hence originates in the density profile of the progenitor~\cite{Nomoto84,Nomoto87}, we assume that the composition of the envelope is entirely helium until the hydrogen burning shell is reached at $r=1090\,\rm km$, and pure hydrogen outside of this point.  In this calculation we neglect direction-changing neutrino-electron scattering.  While the cross sections for neutirno-electron scattering are comparable to the neutrino-neutral current scattering cross sections for helium and free nucleons, the low mass of electrons relative to the energy of neutrinos in the neutronization neutrino burst produces scattering that is strongly forward peaked.  For this reason, neutrinos that have undergone electron scattering do not contribute significantly to either the halo neutrino population or potential.

In keeping with our previously stated $1\,\%$ criterion~\cite{Cherry:2012uq}, we find that outside a radius of $850\,\rm km$ the neutrinos scattered onto inward trajectories do not contribute more than $1\,\%$ to the magnitude of the neutrino self-coupling potential.  We use this result to define a \lq\lq halosphere\rq\rq , at the radius $R_{\rm H} = 850\,\rm km$ for this model, which is the surface outside of which the propagation of halo neutrinos may be taken to be in the outward direction without impacting the dynamics of flavor transformation.  %{\it insert statement about the results of changing} $R_{\rm H}$ {\it here}

A second requirement that must be met is that we must make a physically motivated choice for the initial flavor states of the neutrinos that have been scattered into the halo population.  As mentioned in Section I, multi-angle suppression prevents the onset of neutrino flavor transformation.  We find that the neutrino driven multi-angle suppression alone is sufficient to suppress collective oscillation out to a radius of $r=1100\,\rm km$, and that this figure is little changed by the addition of the halo to our calculations.  Therefore, no neutrinos within this radius will have had the opportunity to engage in collective flavor oscillation.  Because this radius is outside the halosphere surface, all neutrinos scattering into the halo within the halosphere will be in their original flavor state.  Furthermore, because the halo is populated by neutral current processes that are flavor blind, the halo neutrinos will not change flavor after scattering.  This uniquely determines the flavor states of all neutrinos, whether emitted directly from the core or scattered in the halo, at the surface of the halosphere.

\begin{figure}
\centering
\includegraphics[scale=.70]{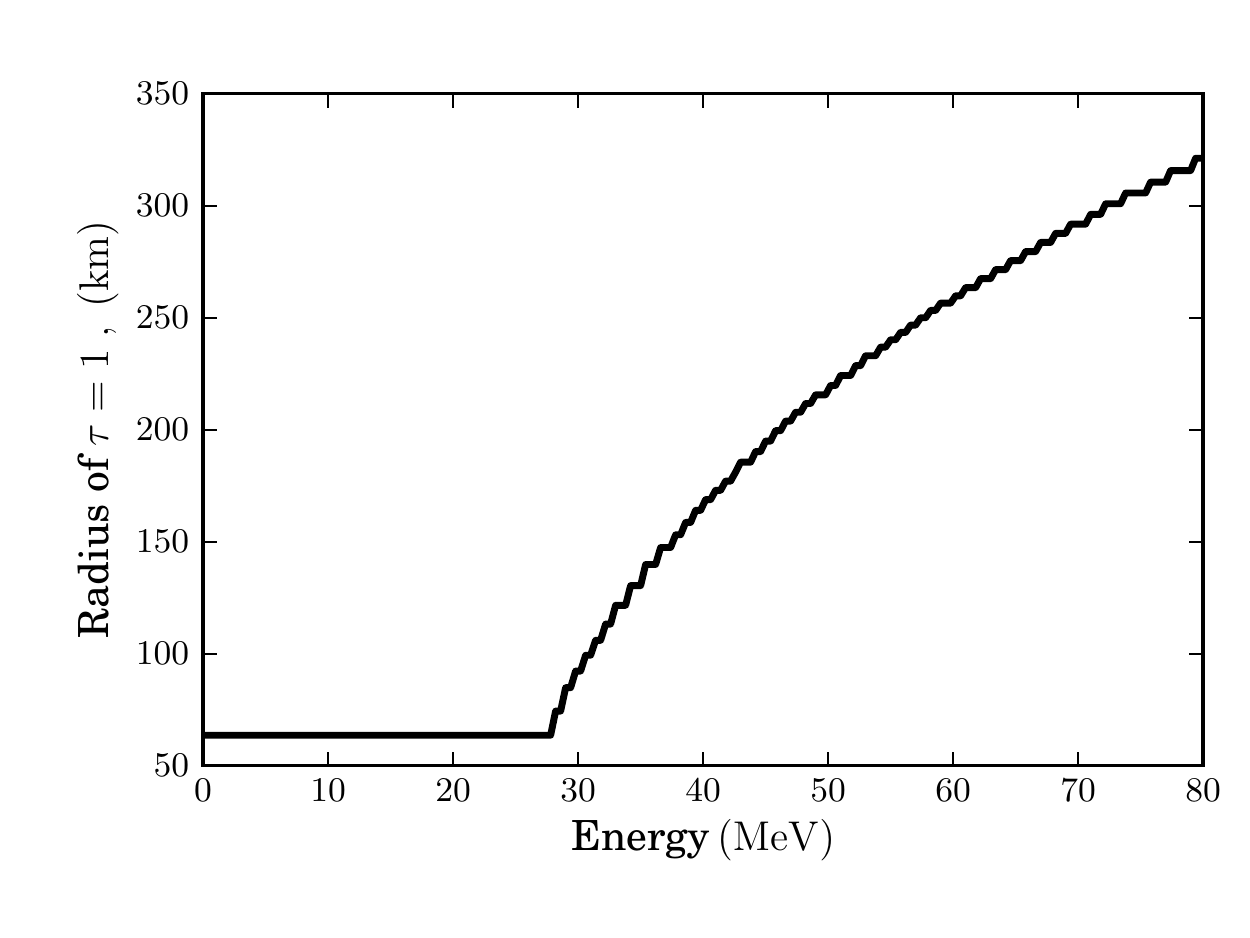}
\caption{The neutrinosphere radius, as defined by the surface of optical depth $\tau = 1$, shown as a function of neutrino energy.}
\label{fig:Nsphere}
\end{figure}

In our previous papers on this subject we have used the convention that all neutrinos are emitted isotropically from a hard spherical shell called the neutrinosphere (which is just above the surface of the proto-neutron star).  A primary criticism of this picture has been that the radius of the neutrinosphere surface itself is dependent on neutrino energy, particularly for high energy neutrinos, and the emission region is extended for these neutrinos.  Adapting our calculation to accommodate differing neutrino emission spectra along different emission trajectories has afforded us the opportunity to rectify this shortcoming.  

A neutrinosphere surface can be defined where the optical depth against scattering is $\tau=1$.  For a typical neutrino (average energy) with our density profile this criterion gives a radius $R_{\nu}=60\,\rm km$.  Each neutrino energy can be assigned an appropriate \lq\lq neutrinosphere\rq\rq\ where the optical depth is unity.  Shown in Figure~\ref{fig:Nsphere} is the computed radius of each $\tau=1$ neutrinosphere surface for neutrinos of a given energy.

\begin{figure}
\centering
\includegraphics[scale=.24]{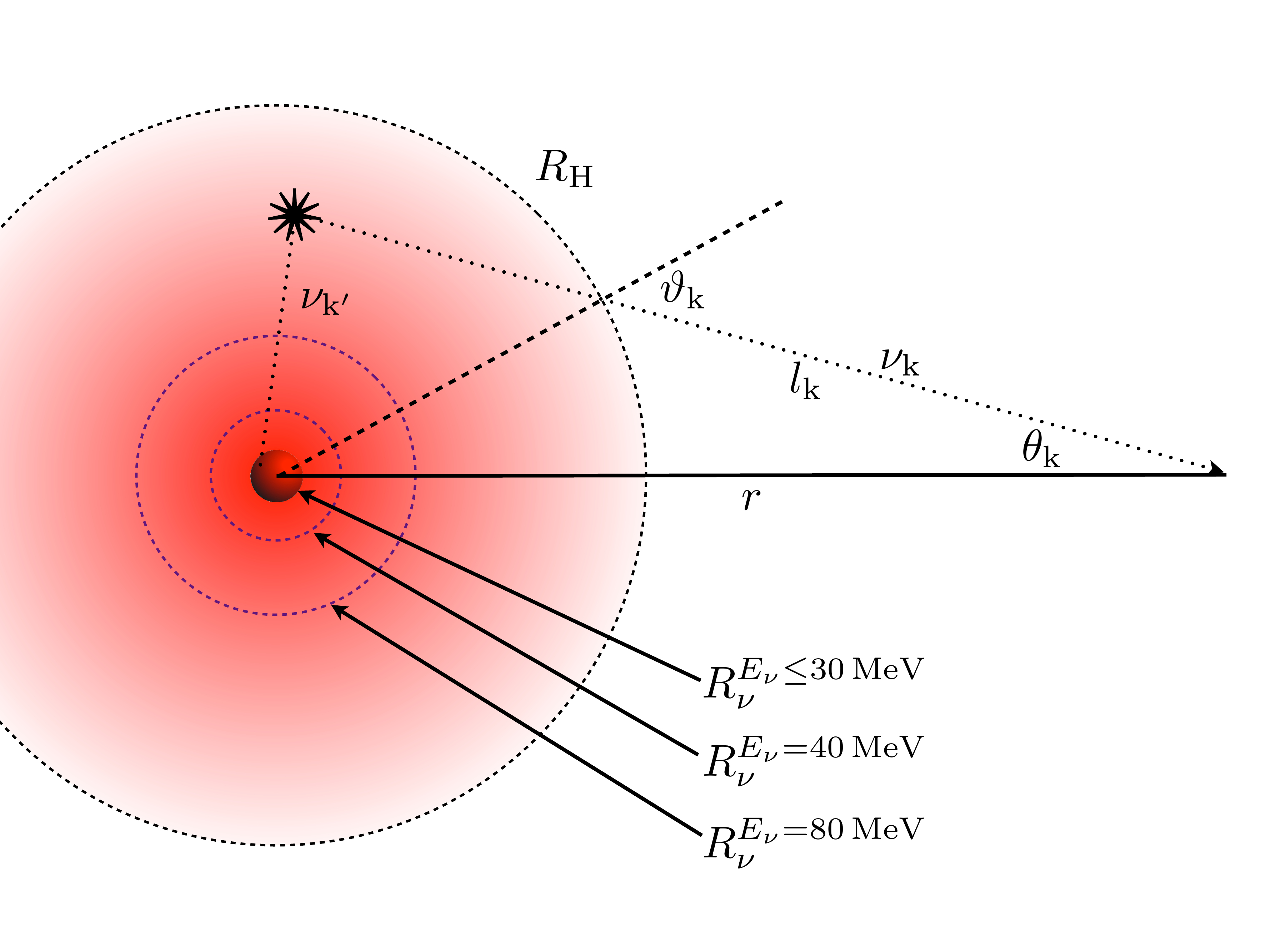}
\caption{Initially neutrinos are emitted isotropically from energy-determined neutrinosphere surfaces, $R^{E_{\nu}}_\nu$.  After emission, some neutrinos are scattered at wide angles into the neutrino halo (e.g. $\nu_{\rm k^\prime}/\nu_{\rm k}$ scattered to location at radius $r$ and angle of incidence $\theta_{\rm k}$), while the remainder continue on unimpeded to the halosphere surface at radius $R_{\rm H}$.  The trajectories of all neutrinos emerging from the halosphere surface are characterized by the angle $\vartheta_{\rm k}$ they make relative to the outward unit normal on this surface and their propagation distance $l_{\rm k}$ from the halosphere surface.}
\label{fig:Hsphere}
\end{figure}

For each neutrino energy we compute the neutral current scattering into the neutrino halo at the surface of the appropriate, energy dependent, neutrinosphere.  This means that the halo region itself is not sharply defined, as low energy neutrinos are scattering into the halo at radii where high energy neutrinos are still inside of their respective neutrinospheres.  The intensity of neutrino emission on the surface of the halosphere is specified by the neutrinos emerging through the surface  along a given trajectory with angle $\vartheta_{\rm k}$ relative to the outward unit normal.  The ensemble of neutrinos at the halosphere surface is populated by emission from the neutrinospheres as well as from the neutrino halo.  Figure~\ref{fig:Hsphere} shows a cartoon representation of the physical set up of the initial conditions where $\vartheta_{\rm k}$ is the neutrino emission angle relative to the surface of the halosphere, $r$ is the distance from the center of the supernova, $\theta_{\rm k}$ is the angle of intersection of the neutrino trajectory with the outward unit normal, and $l_{\rm k}$ is the propagation distance of $\nu_{\rm k}$ from the surface of the halosphere to the location at radius $r$.

\begin{figure}
\centering
\includegraphics[scale=.70]{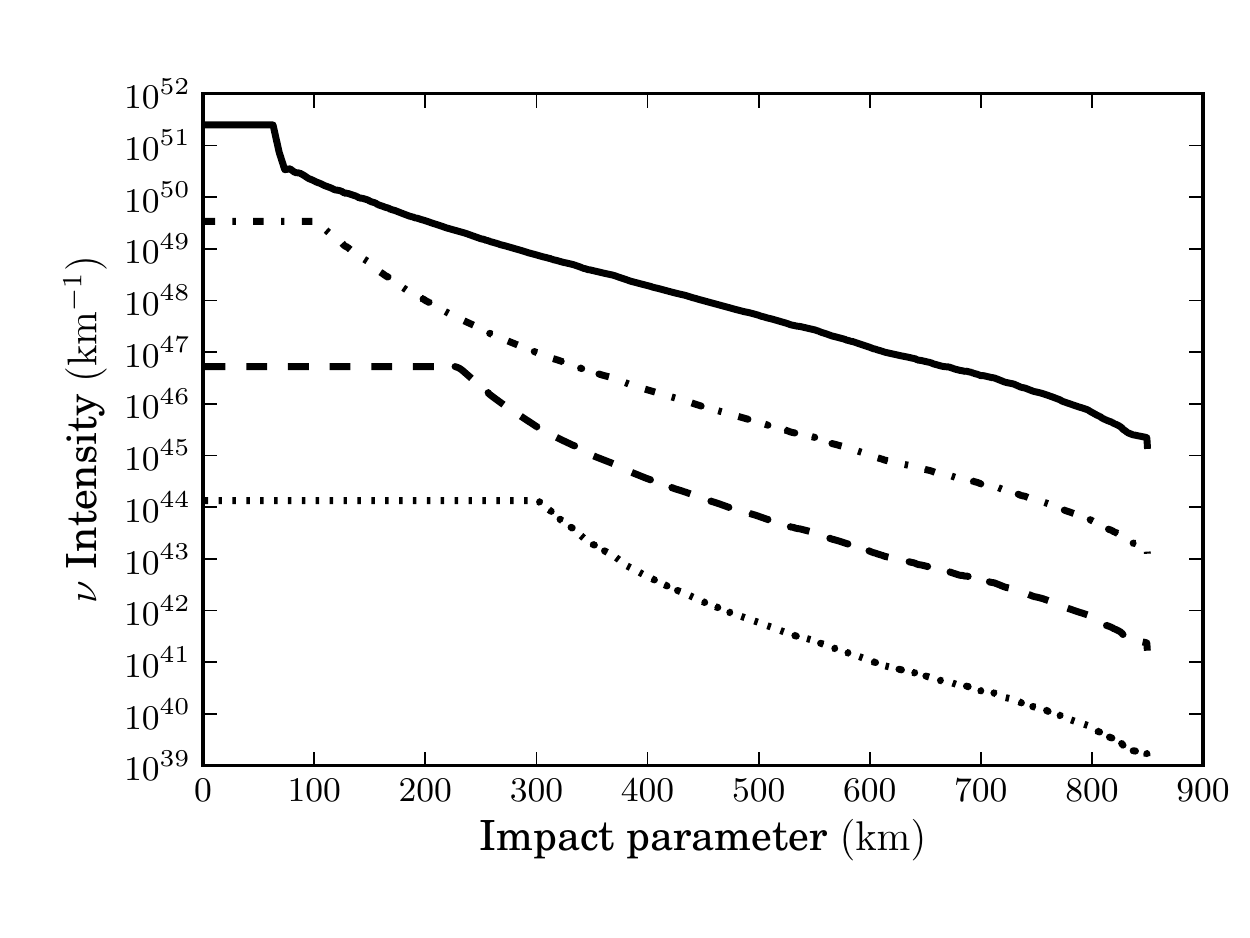}
\caption{The neutrino emission intensity as a function of impact parameter at the halosphere surface.  Line style indicates different neutrino energies: $10\,\rm MeV$ (solid), $30\,\rm MeV$ (dash-dotted), $50\,\rm MeV$ (dashed), $70\,\rm MeV$ (dotted).}
\label{fig:Ivb}
\end{figure}

This approach carries the explicit assumption that there is a clear break between the multiple scattering regime, where the optical depth of neutrinos is $\tau > 1$, and the single scattering regime, where $\tau < 1$, at the (energy dependent) neutrinosphere surface.  We take the emission of neutrinos of a given energy to be isotropic at the neutrinosphere surface (which is the limit in the case of multiple neutrino scattering), and employ only single scattering to compute the angular dependence of the halo neutrino population outside the appropriate neutrinosphere.  While this approach is somewhat crude, it produces results that are in remarkably good agreement with sophisticated models of neutrino transport in supernovae~\cite{Bruenn:2010qy,Muller:2010kx,Brandt:2011lr,Sarikas:2012qy}.  

The intensity of the neutrino emission at the halosphere surface is shown in Figure~\ref{fig:Ivb}.  Note that because of the existence of multiple neutrinospheres and the diffuse scattering of halo neutrinos, we have chosen to parameterize the individual neutrino trajectories by their impact parameter, $b = R_{\rm H} \sin{\vartheta_{\rm k}}$, relative to the center of the proto-neutron star.  For a given neutrino energy $E_\nu$, the neutrino emission \lq\lq intensity\rq\rq , $I_\nu\left( E_\nu\right)$, shown in Figure~\ref{fig:Ivb} is defined in this case to be related to the neutrino number density, $n_\nu\left( E_\nu\right)$, outside the halo sphere surface by the relation,
\begin{equation}
n_\nu\left( E_\nu\right) = \int_{0}^{\theta_{\rm max}} I_\nu\left( E_\nu\right) \frac{4\pi \cos\vartheta_{\rm k}}{l_{\rm k}^2} d\cos\theta_{\rm k}\, .
\end{equation}

\section{Results}

\begin{figure*}
\centering
\includegraphics[scale=.70]{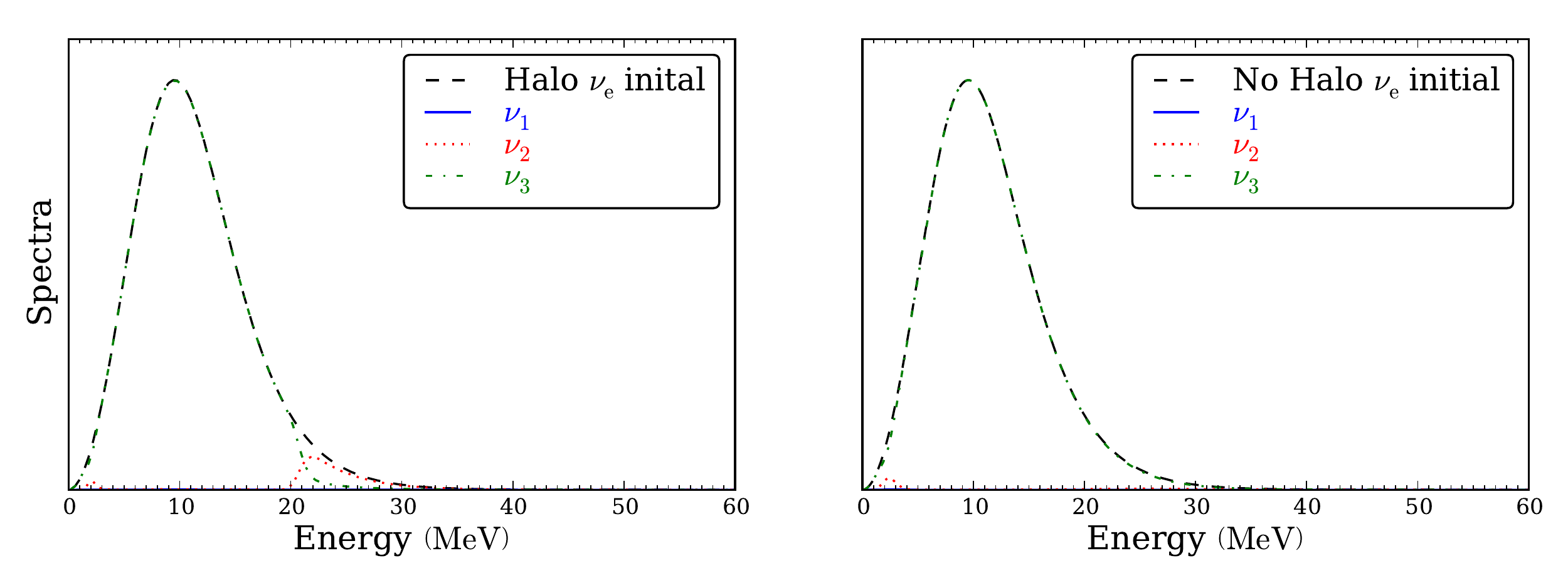}
\caption{A comparison of the emission angle averaged results of flavor transformation calculations with the halo neutrinos included and with halo scattering neglected.  Left panel: the calculation including the halo, mass basis (key top right, inset) neutrino energy distribution functions versus neutrino energy.  The dashed curve gives the initial $\nu$ energy spectrum.  Right panel:  the calculation neglecting halo scattering, mass basis (key top right, inset) neutrino energy distribution functions versus neutrino energy.  The dashed curve gives the initial $\nu$ energy spectrum.  Both panels show the final state of neutrino flavor transformation at a radius of $r=12000\,{\rm km}$.}
\label{fig:SpecComp}
\end{figure*}

\begin{figure}
\centering
\includegraphics[scale=.63]{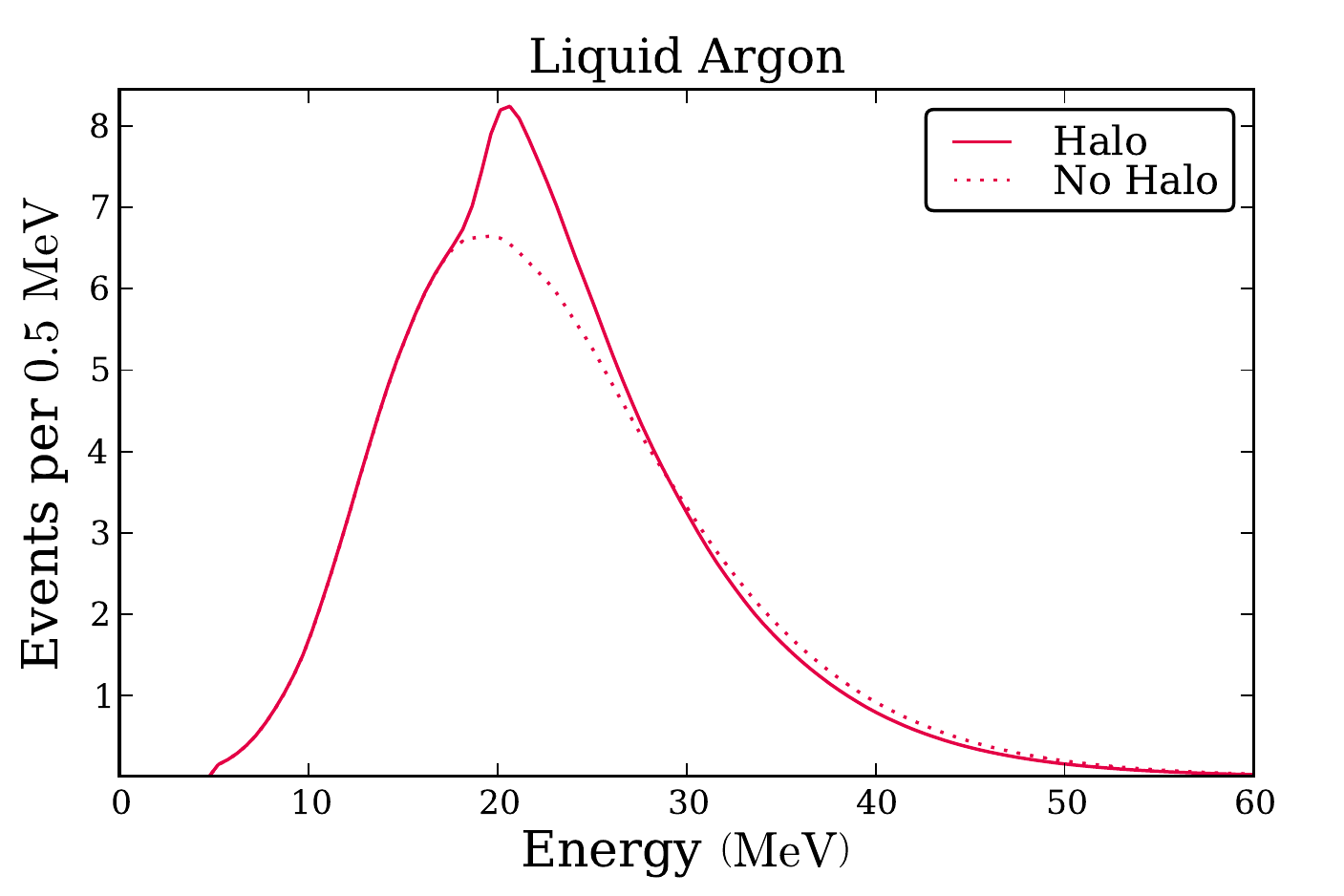}
\caption{A comparison of the modeled event rate for detected $\nu_{\rm e}$ captures in a $17\,\rm kt$ liquid Argon detector between calculations with and without the scattered neutrino halo.  The spectral distortions created by the halo produce a clear swap signature between $20 - 30\,\rm MeV$, which constitute $\sim 15$ additional $\nu_{\rm e}$ events in this $20\,\rm ms$ time slice of the supernova signal.}
\label{fig:SigComp}
\end{figure}

Our results show that there are both quantitative and qualitative changes in the structure of the neutrino flavor transformation in the presence neutrino halo. We consider two different time slices of the neutronization burst, one $7\,\rm ms$ post core bounce at the height of the $\nu_{\rm e}$ luminosity, and another $15\,\rm ms$ post bounce when the fluxes of other species of neutrino have begun to rise appreciably.  All neutrino emission parameters are taken directly from Reference~\cite{Fischer:2010lq}.

The salient question raised by the addition of halo scattering to our flavor transformation calculations is whether or not this effect has any observable consequence.  To address this, we have taken the fluxes of neutrinos generated by our flavor transformation calculations for both the halo and no halo scattering cases and used the {\it SNOwGLoBES} software package~\cite{Scholberg:2011kj} to model the detected signal corresponding to the two spectra seen in Figure~\ref{fig:SpecComp}.  The primary signal for the neutronization burst epoch will be in the $\nu$ sector, so we have chosen to use a liquid Argon model detector that is most sensitive to the $\nu_{\rm e}$ flux via charged current capture~\cite{Scholberg:2012kj}.  The emission angle averaged results of our calculations for the $15\,\rm ms$ post bounce time slice are shown in Figure~\ref{fig:SpecComp}.  The results in Figure~\ref{fig:SpecComp} show that the addition of the halo scattering has produced a swapped population of mass state 2 neutrinos above an energy of $20\,\rm MeV$.  Because mass state $2$ has a larger electron flavor component than mass state $3$, the swapping of neutrinos into mass state $2$ shown if Figure~\ref{fig:SpecComp} is expected to produce a signal which should be observable.  Shown in Figure~\ref{fig:SigComp} is the difference between the modeled detector signals with and without halo neutrino scattering, for the portion of the neutronization burst where the emitted fluxes are similar to what was used in the flavor transformation calculations shown in Figure~\ref{fig:SpecComp}, a window of $\sim 20\,\rm ms$.  Figure~\ref{fig:SigComp} demonstrates that the effect of the halo scattering has been to produce a clear swap feature that is detectable in the received supernova neutronization burst signal.

\begin{figure*}
\centering
\includegraphics[scale=.70]{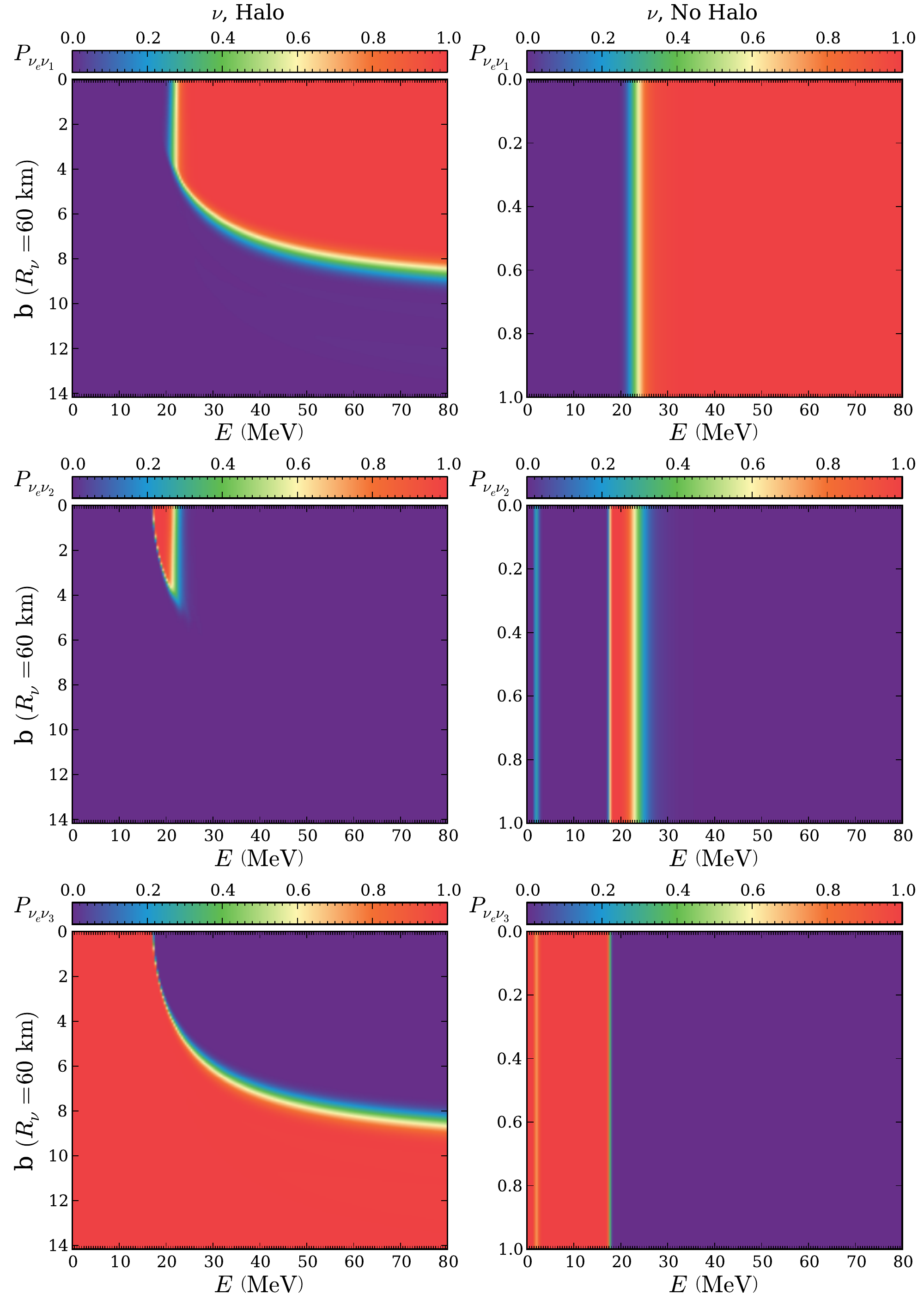}
\caption{Results of flavor transformation calculations with the halo neutrinos included.  Initial emission spectra are taken from~\cite{Fischer:2010lq}, $7\,\rm ms$ post core bounce.  Left panels: electron neutrino occupation probability $P_{\nu_{\rm e}\nu_{\rm x}}$ (color/shading key at top of panel), where $x=1,\, 2,\, 3$ is the neutrino mass eigenstate, shown as a function of impact parameter, $b$ in units of neutrinosphere radius $R_\nu=60\,\rm km$, and neutrino energy, $E$ in MeV, plotted at a radius of $r=12000\,{\rm km}$.  Right panels:  electron anti-neutrino occupation probability $P_{\bar\nu_{\rm e}\bar\nu_{\rm x}}$ (color/shading key at top of panel), where $x=1,\, 2,\, 3$ is the anti-neutrino mass eigenstate, shown as a function of impact parameter, $b = R_{\rm H}\sin\vartheta_{\rm k}$, and neutrino energy, $E$ in MeV, plotted at a radius of $r=12000\,{\rm km}$.}
\label{fig:MB7}
\end{figure*}

\begin{figure*}
\centering
\includegraphics[scale=.70]{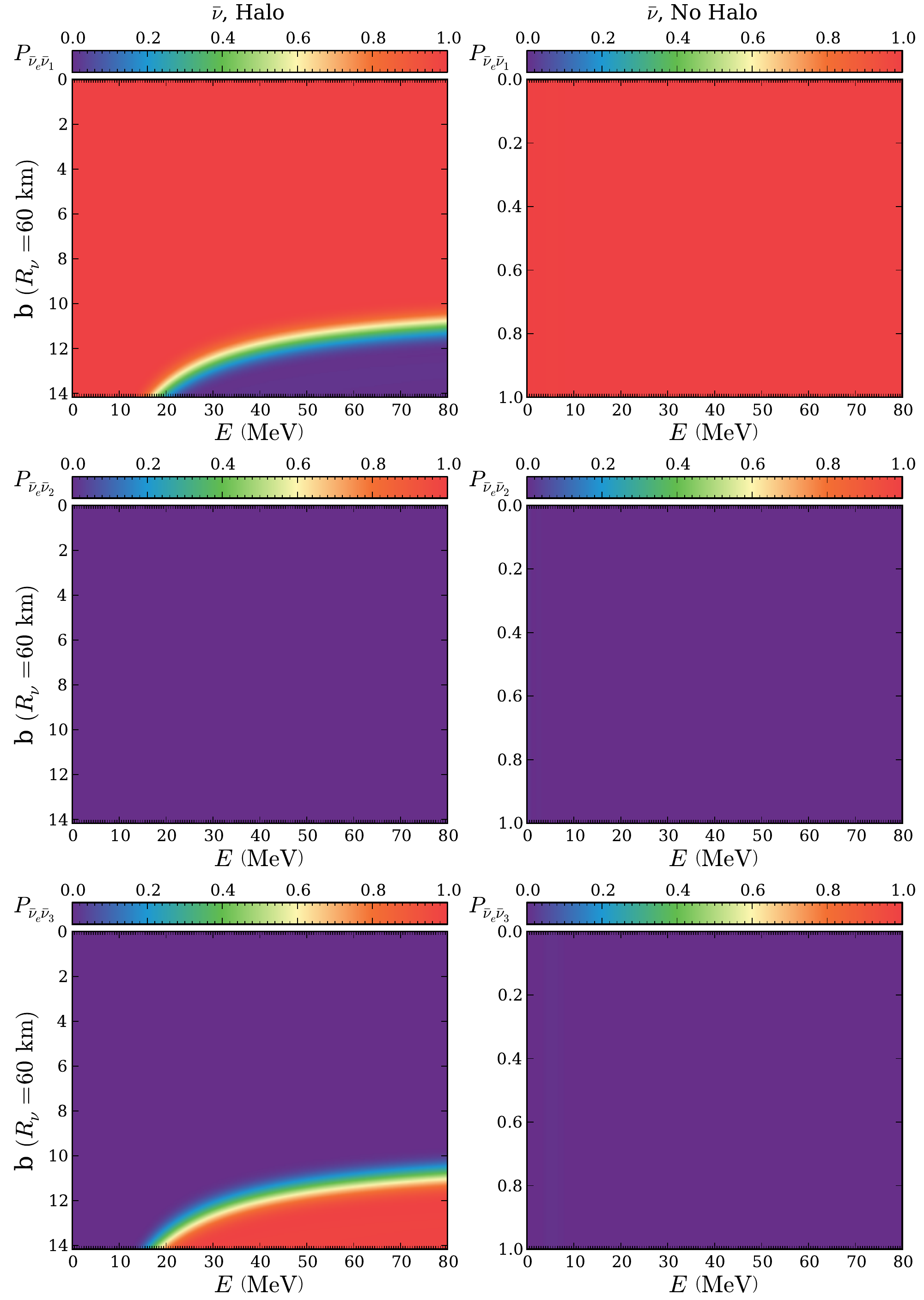}
\caption{Same as Figure~\ref{fig:MB7} for $\bar\nu_{\rm e}$'s.}%Results of flavor transformation calculations where all neutrinos are retained in the neutrinosphere emission, and not allowed to scatter into the halo.  Initial emission spectra are taken from~\cite{Fischer:2010lq}, $7\,\rm ms$ post core bounce.  Left panels: electron neutrino occupation probability $P_{\nu_{\rm e}\nu_{\rm x}}$ (color/shading key at top of panel), where $x=1,\, 2,\, 3$ is the neutrino mass eigenstate, shown as a function of impact parameter, $b$ in units of neutrinosphere radius $R_\nu=60\,\rm km$, and neutrino energy, $E$ in MeV, plotted at a radius of $r=12000\,{\rm km}$.  Right panels:  electron anti-neutrino occupation probability $P_{\bar\nu_{\rm e}\bar\nu_{\rm x}}$ (color/shading key at top of panel), where $x=1,\, 2,\, 3$ is the anti-neutrino mass eigenstate, shown as a function of impact parameter, $b = R_{\rm H}\sin\vartheta_{\rm k}$, and neutrino energy, $E$ in MeV, plotted at a radius of $r=12000\,{\rm km}$.}
\label{fig:MB7B}
\end{figure*}

\begin{figure*}
\centering
\includegraphics[scale=.70]{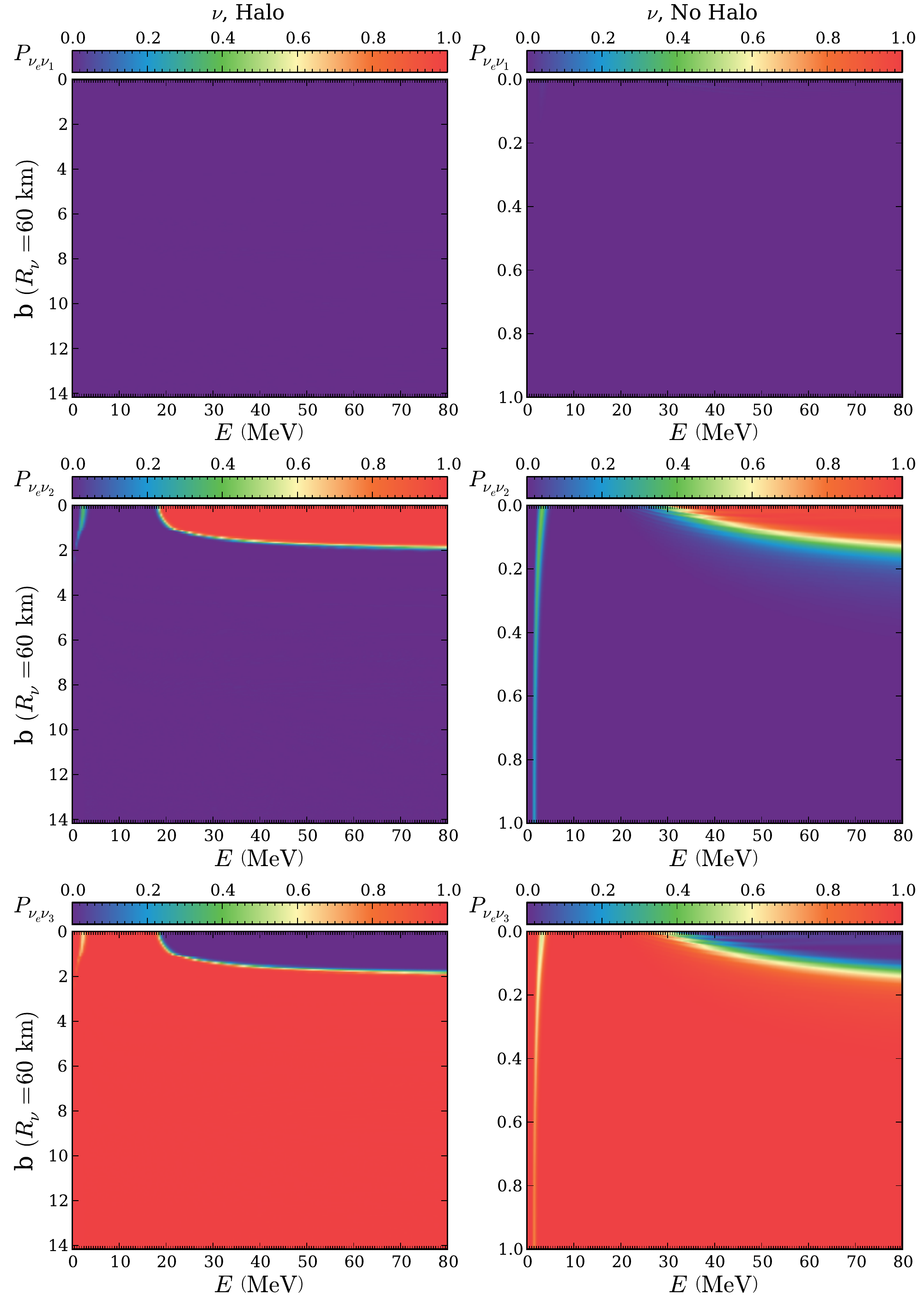}
\caption{Same as Figure~\ref{fig:MB15} for $\nu_{\rm e}$'s during the $15\,\rm ms$ time snapshot.}%Results of flavor transformation calculations where all neutrinos are retained in the neutrinosphere emission, and not allowed to scatter into the halo.  Initial emission spectra are taken from~\cite{Fischer:2010lq}, $15\,\rm ms$ post core bounce.  Left panels: electron neutrino occupation probability $P_{\nu_{\rm e}\nu_{\rm x}}$ (color/shading key at top of panel), where $x=1,\, 2,\, 3$ is the neutrino mass eigenstate, shown as a function of impact parameter, $b$ in units of neutrinosphere radius $R_\nu=60\,\rm km$, and neutrino energy, $E$ in MeV, plotted at a radius of $r=12000\,{\rm km}$.  Right panels:  electron anti-neutrino occupation probability $P_{\bar\nu_{\rm e}\bar\nu_{\rm x}}$ (color/shading key at top of panel), where $x=1,\, 2,\, 3$ is the anti-neutrino mass eigenstate, shown as a function of impact parameter, $b = R_{\rm H}\sin\vartheta_{\rm k}$, and neutrino energy, $E$ in MeV, plotted at a radius of $r=12000\,{\rm km}$.}
\label{fig:MB15}
\end{figure*}

\begin{figure*}
\centering
\includegraphics[scale=.70]{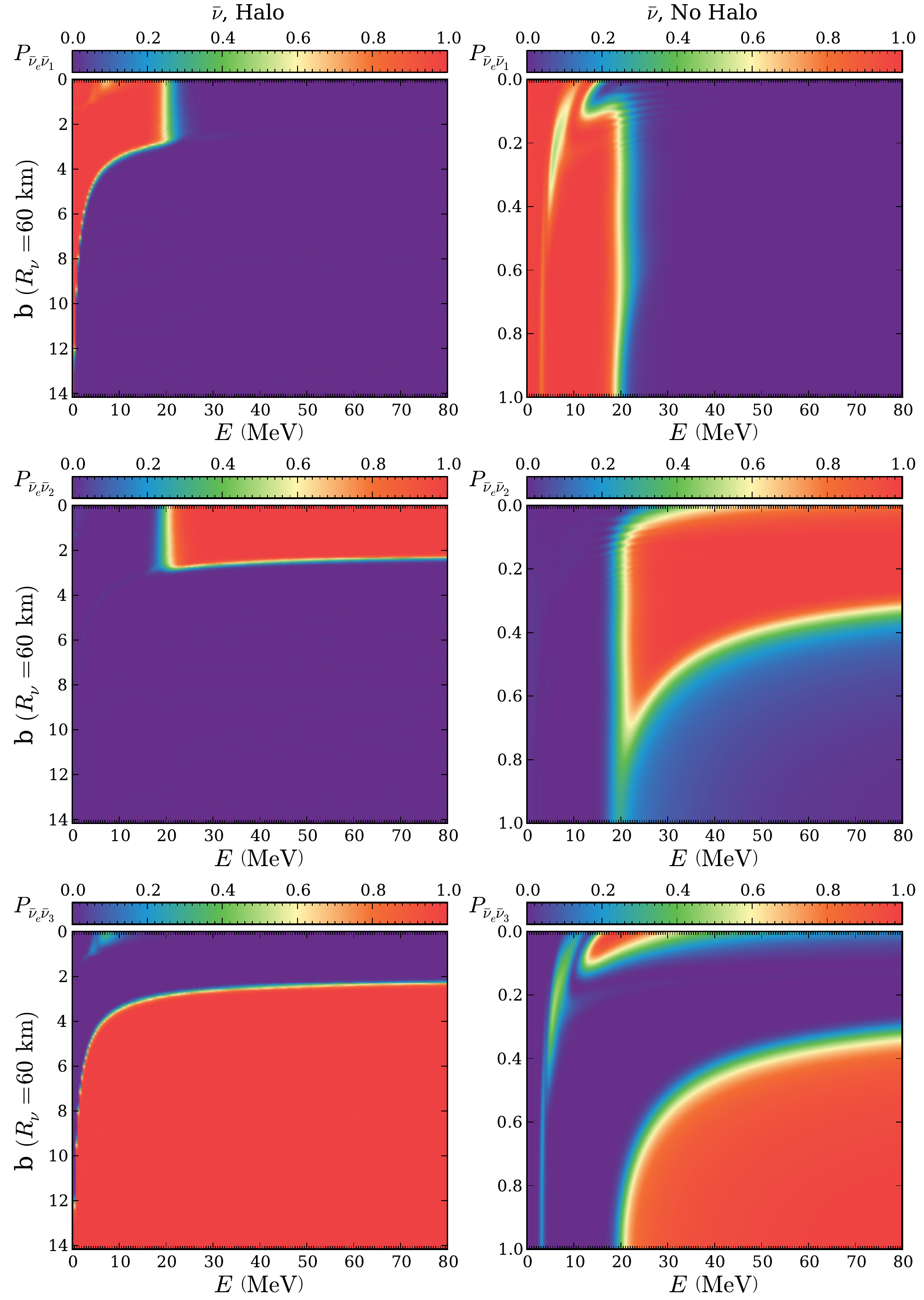}
\caption{Same as Figure~\ref{fig:MB15} for $\bar\nu_{\rm e}$'s during the $15\,\rm ms$ time snapshot.}%Results of flavor transformation calculations with the halo neutrinos included.  Initial emission spectra are taken from~\cite{Fischer:2010lq}, $15\,\rm ms$ post core bounce.  Left panels: electron neutrino occupation probability $P_{\nu_{\rm e}\nu_{\rm x}}$ (color/shading key at top of panel), where $x=1,\, 2,\, 3$ is the neutrino mass eigenstate, shown as a function of impact parameter, $b$ in units of neutrinosphere radius $R_\nu=60\,\rm km$, and neutrino energy, $E$ in MeV, plotted at a radius of $r=12000\,{\rm km}$.  Right panels:  electron anti-neutrino occupation probability $P_{\bar\nu_{\rm e}\bar\nu_{\rm x}}$ (color/shading key at top of panel), where $x=1,\, 2,\, 3$ is the anti-neutrino mass eigenstate, shown as a function of impact parameter, $b = R_{\rm H}\sin\vartheta_{\rm k}$, and neutrino energy, $E$ in MeV, plotted at a radius of $r=12000\,{\rm km}$.}
\label{fig:MB15B}
\end{figure*}

Figures 7 - 10 show in detail the results of our flavor transformation calculations.  These figure show the probabilities for a neutrino or anti-neutrino, initially in the electron flavor state, to occupy each of the neutrino mass basis states at large radius, after neutrino flavor transformation is complete.   We define this probability to be:
\begin{equation}
P_{\nu_{\rm i}\nu_{\rm a}} \equiv \vert\langle \Psi_{\nu_{\rm i}}\left( r_{\rm initial} \right) \vert 
\Psi_{\nu_{\rm a}}\left( r_{\rm final} \right) \rangle \vert^2 \, .
\end{equation}
Here $\vert\Psi_{\nu_{\rm i}}\left( r_{\rm initial} \right) \rangle$ is the wave function for a neutrino of flavor $i= e,\,\mu,\,\tau$ at the radius $r_{\rm initial}$ where the neutrino flavor states are initialized, and $\vert \Psi_{\nu_{\rm a}}\left( r_{\rm final} \right) \rangle$ is the wave function for a neutrino mass state $a = 1,\, 2,\, 3$ at the radius $r_{\rm final}$ where the calculation ends.  Figure~\ref{fig:MB7} shows the mass state occupation probabilities for electron neutrinos during the $7\,\rm ms$ time slice, comparing the cases where the initial states of all neutrinos are prepared in with and without scattering into the halo population.  Figure~\ref{fig:MB7B} shows the mass state occupation probabilities for electron anti-neutrinos during the $7\,\rm ms$ time slice, comparing the cases where the initial states of all neutrinos are prepared in with and without scattering into the halo population.   
Similarly, Figures~\ref{fig:MB15} and ~\ref{fig:MB15B} display the comparison of the Halo and No Halo cases for the $\nu_{\rm e}/\bar\nu_{\rm e}$ mass state occupation probabilities, respectively, for the $15\,\rm ms$ time snapshot.

For the cases where no halo scattering is included, the emission from the surface of the neutrinosphere is isotropic, and the trajectory bins that lie in the halo region are unpopulated.  For each time snapshot the Halo vs. No Halo calculations employ identical energy spectra for all neutrino flavors, and identical binning schemes for both emission angle and neutrino energy.  Furthermore, the emission trajectory binning for both calculations was chosen so that there would be an exact match to the bins on the surface of the neutrinosphere employed in previous calculations~\cite{Cherry:2012lr}, with an equal number of additional trajectory bins added to accommodate the halo neutrinos.

The most noticeable feature of Figures~\ref{fig:MB7} and~\ref{fig:MB7B} is that the mass state $3/2$ swap is emission angle independent when neglecting the halo effect, and very much dependent on the neutrino trajectory when the halo is included.  This is a surprising result, as the analytic work which has studied the formation of these flavor swaps does not find any dependence of the swap energy on the emission angle.  Fascinatingly, with the halo the swap interface is pushed through $E_{\nu} = \infty$ at large emission angles, and reappears in the anti-neutrino sector, creating a swap in the anti-neutrinos which is not present in the absence of the halo.  Even more interesting is that unlike the  mass state $3/2$ swap, the swap between mass state $2/1$ exhibits an apparent lack of emission trajectory dependence.

For the Figures~\ref{fig:MB15} and~\ref{fig:MB15B} the disparity of between number fluxes of $\nu_{\rm e}$ and $\bar\nu_{\rm e}$ has diminished after the burst earlier reached peak luminosity,which is typical of the neutronization burst of core collapse models from which these spectra are drawn~\cite{Fischer:2010lq}.  As can be seen in Figure~\ref{fig:MB15B}, this shift in the spectral properties of the emission has moved the bulk of the spectral swaps into the $\bar\nu$ sector when scattering into the halo is neglected.  However, when the effect of the halo is included the spectral distortion it produces shifts the mass state $3/2$ swap back into the $\nu$ sector.  This creates a population of swapped mass state $2$ neutrinos above $\sim 20\,\rm MeV$, which was not present when the calculation was performed in the absence of the halo.  As we discussed earlier in this section, this distortion of the spectral swap created by the halo scattering produces a clearly detectable swap signal in the $\nu$ sector for this time snapshot.

\section{Theory}

\begin{figure*}
\centering
\includegraphics[scale=.70]{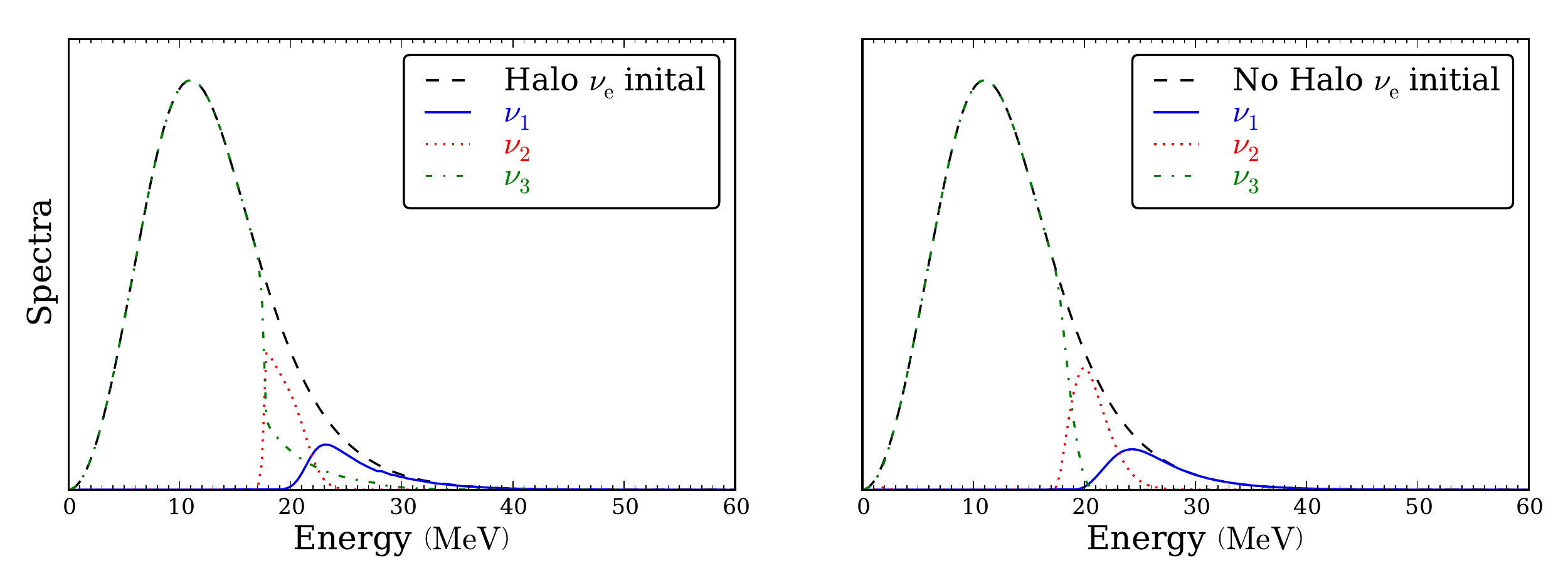}
\caption{Emission angle averaged results of flavor transformation calculations at $7\,\rm ms$ post core bounce comparing calculations with and without halo neutrino scattering.  Left panel: mass basis (key top right, inset) neutrino energy distribution functions versus neutrino energy when halo scattering is included in the calculation.  The dashed curve gives the initial $\nu_{\rm e}$ energy spectrum.  Right panel:  mass basis (key top right, inset) neutrino energy distribution functions versus neutrino energy when halo scattering is excluded from the calculation.  The dashed curve gives the initial $\bar\nu_{\rm e}$ energy spectrum.  Both panels show the final state of neutrino flavor transformation at a radius of $r=12000\,{\rm km}$.}
\label{fig:MBH}
\end{figure*}

\begin{figure*}
\centering
\includegraphics[scale=.70]{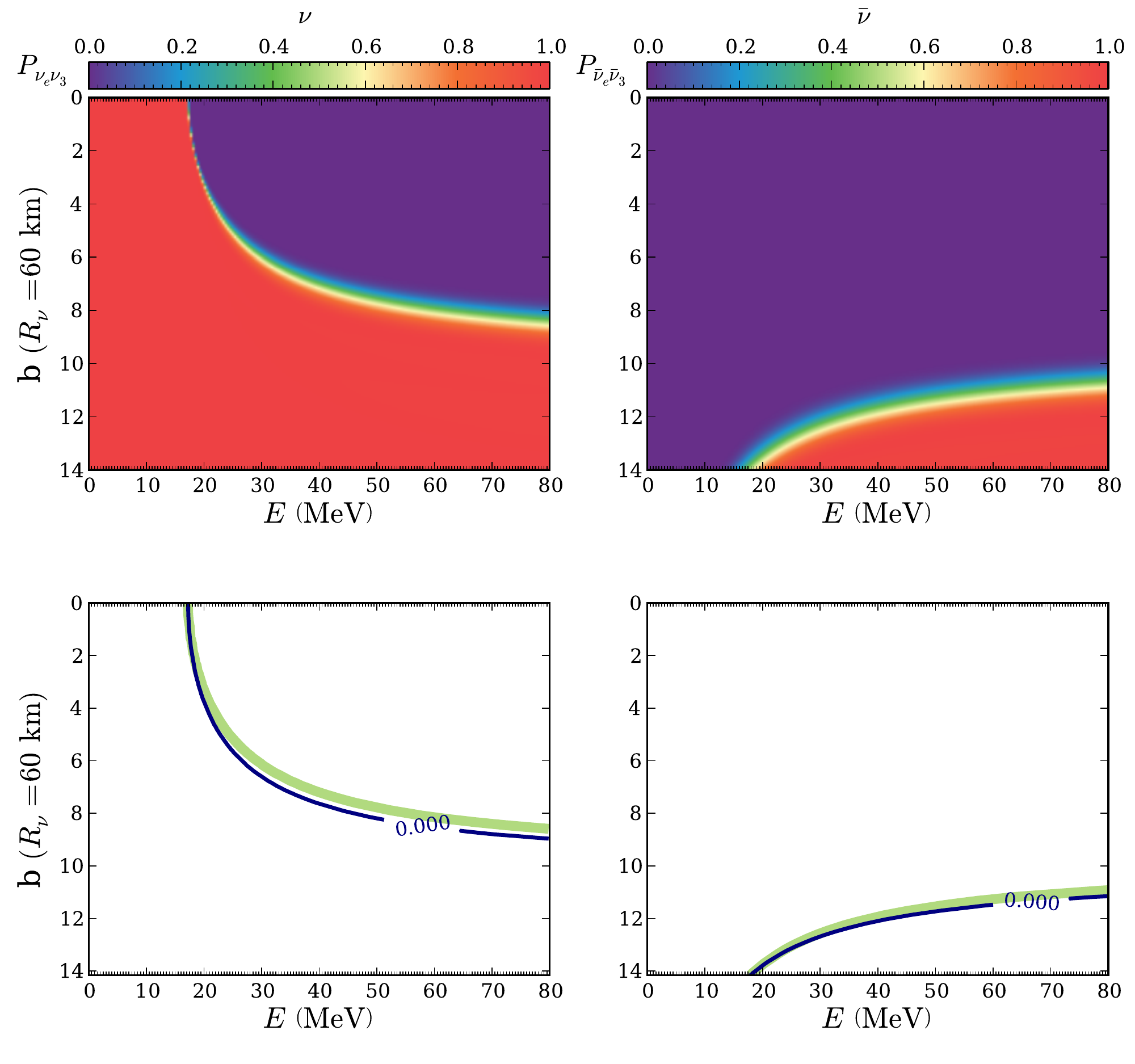}
\caption{Top panels: the structure of the mass state $3/2$ swap in the neutrino (left) and anti-neutrino (right) sectors, displayed in terms of the mass state 3 occupation probability for the $7\,\rm ms$ post bounce calculation  (notation and axes as in Figure~\ref{fig:MB7}).  Bottom panels: shown in black is the contour which satisfies the emission trajectory dependent swap criterion of Equation~\ref{newswap} for neutrinos (left) and anti-neutrinos (right).  The thicker, lighter color contour shows the locations where a $\nu_{\rm e}$ or $\bar\nu_{\rm e}$ has a $50\,\%$ probability to occupy mass state 3.  The contours displayed are selected at a radius where $\vert \hat{H}_{\nu\nu}\vert \sim \omega_{\rm pr}$ is satisfied for a given impact parameter, $b$.}
\label{fig:SwapSurf}
\end{figure*}

\begin{figure*}
\centering
\includegraphics[scale=.70]{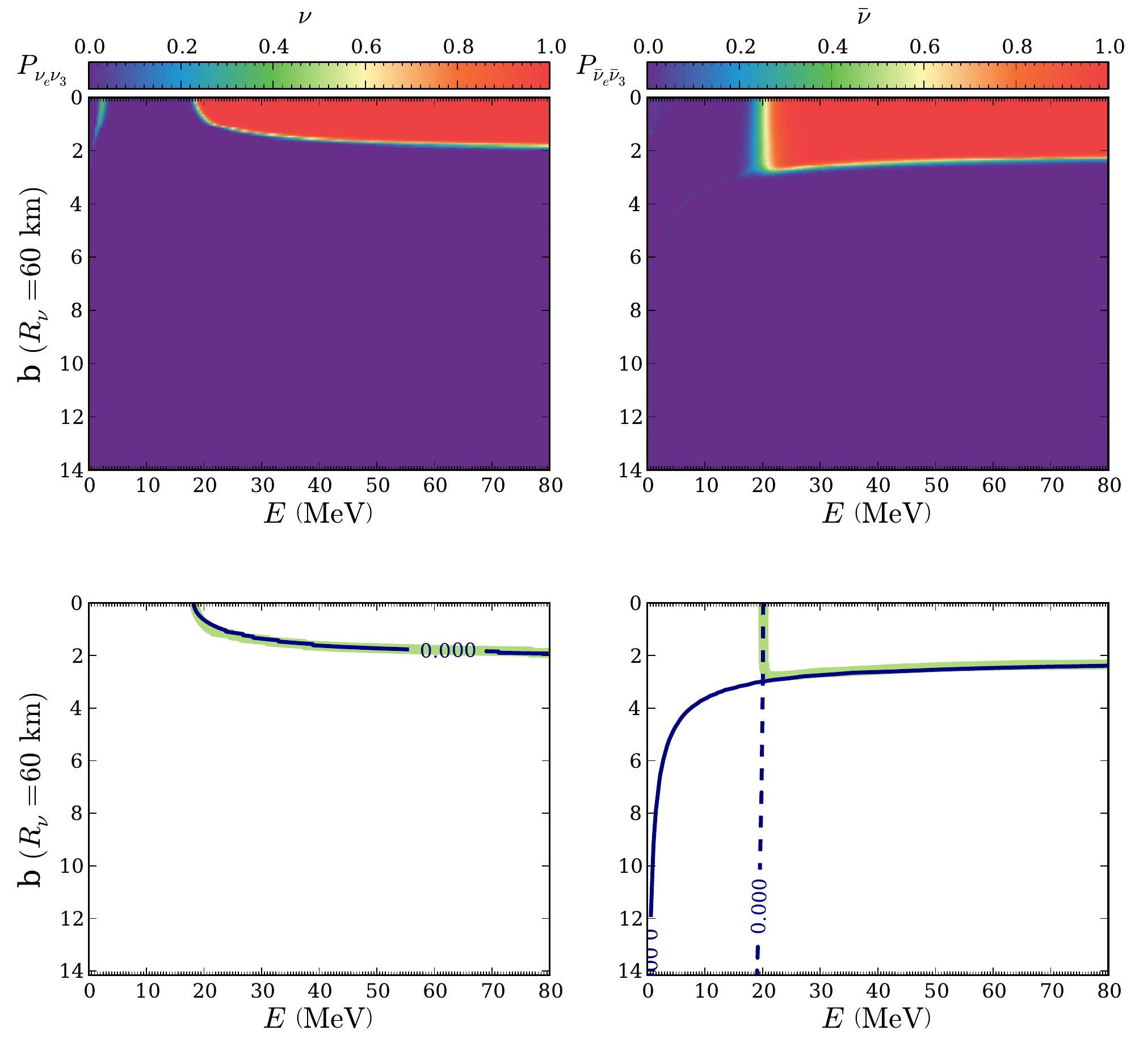}
\caption{Top panels: the structure of the mass state $2$ occupation probability in the neutrino (left) and anti-neutrino (right) sectors for the $15\,\rm ms$ post bounce calculation  (notation and axes as in Figure~\ref{fig:MB7}).  Bottom panels: shown in solid black is the contour which satisfies the emission trajectory dependent swap criterion of Equation~\ref{newswap} for the mass state $3/2$ swap for neutrinos (left) and anti-neutrinos (right).  Shown in dashed black is the contour which satisfies the emission trajectory dependent swap criterion of Equation~\ref{newswap} for the mass state $2/1$ swap for anti-neutrinos (right).  The thicker, lighter color contour shows the locations where a $\nu_{\rm e}/\bar\nu_{\rm e}$ has a $50\,\%$ probability to occupy mass state 2.  The mass state $3/2$ contour and the mass state $2/1$ contour are selected at differing radius so that for both contours the condition $\vert \hat{H}_{\nu\nu}\vert \sim \omega_{\rm pr}$ is satisfied individually for each combination of precession frequency, $\omega_{\rm pr}$, and impact parameter, $b$.}
\label{fig:SwapSurf2}
\end{figure*}

%\begin{figure*}
%\centering
%\includegraphics[scale=.73]{./MB_spectra_test_nm.pdf}
%\caption{Emission angle averaged results of flavor transformation calculations at $7\,\rm ms$ post core bounce where all neutrinos are retained in the neutrinosphere emission, and not allowed to scatter into the halo.  Left panel: mass basis (key top right, inset) neutrino energy distribution functions versus neutrino energy.  The dashed curve gives the initial $\nu_{\rm e}$ energy spectrum.  Right panel:  mass basis (key top right, inset) anti-neutrino energy distribution functions versus neutrino energy.  The dashed curve gives the initial $\bar\nu_{\rm e}$ energy spectrum.  Both panels show the final state of neutrino flavor transformation at a radius of $r=12000\,{\rm km}$.}
%\label{fig:MBNH}
%\end{figure*}
The spectral distortions found in our calculations raise a question:  Do the halo neutrinos, though few in number, nevertheless alter the qualitative and quantitative character of collective neutrino oscillations?  The answer: At $7\,\rm ms$ in our model the halo primarily affects the collective oscillations of neutrinos propagating at large impact parameters; but $8\,\rm ms$ later the halo neutrinos completely re-determine the course of neutrino flavor oscillation for all emission trajectories.  This result underscores the necessity for a self-consistent numerical treatment of this nonlinear system.

The twisting of one of the swap surfaces through the trajectory space has several direct consequences.  The first is the shift in the swap energies.  When the halo effect is included in the $7\,\rm ms$ post bounce case, a high energy tail of $\nu_3$ remains unswapped in the neutrino sector.  Figure~\ref{fig:MBH} shows this feature in the total angle-averaged energy spectra for electron neutrinos projected into the three mass states for our simulation with and without the halo.  The total number of neutrinos in each mass state for both the halo and no-halo cases are nearly identical (there are small differences on the order of $\sim 0.1\,\%$, owing to slight increases in the adiabaticity of flavor evolution when the halo is included).  With the halo the number of neutrinos that remain in mass state $3$ at high energy causes the swap between $\nu_3 / \nu_2$ to form at lower energy.  Consequently, this also lowers the swap energy for mass states $\nu_2 / \nu_1$.

This can be understood simply from the equations of motion.  The collective flavor oscillation which creates the swaps (called the Regular Precession Mode) in this example posseses two conserved constants of the motion, effective lepton numbers for each of the mass-squared splittings~\cite{Duan:2008qy}.  Because the scattering of neutrinos into the halo does not change the spectral shape of the entire ensemble of neutrinos, one might reasonably expect that the conserved lepton numbers that describe the flavor evolution of the neutrinos to remain unchanged by the presence of the halo.  Indeed, this is what is found in our calculations.  Following the convention of Reference~\cite{Duan:2008qy}, the conserved lepton numbers $L_8$ and $L_3$ for the atmospheric and solar mass squared splittings, respectively, are identical for calculations with and without the presence of the halo.  As a consequence of this conservation law, the presence of additional neutrinos in mass state three at high energy necessitates a concomitant reduction in the swap energy seen in Figure~\ref{fig:MB7}.

The conserved lepton numbers are also nearly identical for the calculations performed for the neutrino emission $15\,\rm ms$ post core bounce.  Although the total occupation of mass state 2 in the neutrino sector is manifestly different for the Halo vs. the No Halo case, the contributions of $\nu$'s and $\bar\nu$'s to the magnitude of $L_8$ and $L_3$ have a relative sign difference.  Once the number fluxes of $\bar\nu$ begin to rise during the neutronization burst, the spectral distortions in the $\bar\nu$ sector can significantly alter the $\bar\nu$ contributions to $L_8$ and $L_3$.  To satisfy the conservation of lepton number demanded by the equations of motion, the spectral distortions in the $\nu$ sector must shift an appreciable number neutrinos into the appropriate mass Eigen states to balance the spectral distortions in the $\bar\nu$ sector.

The agreement on the conserved lepton numbers between the Halo and No Halo calculations strongly suggests that the distortion of the swap surface seen in Figures 7 - 10 is due to differences in the geometry and spatial distribution of neutrinos between the halo and no halo calculations, and is not a product of neutrinos following a different equation of motion through flavor space in the presence of the halo.  The Regular Precession Mode is characterized by strong non-linear coupling of neutrino flavor states, leading to all neutrinos in the ensemble oscillating in flavor space with the same frequency, $\omega_{\rm pr}$.  In the single angle approximation all neutrinos either align or anti-align with a mass eigenstate as the neutrino self-coupling potential decreases, and $\omega_{\rm pr} = \omega_{\rm swap} = \Delta m^2 / 2 E_{\rm swap}$.  The sense of the swap depends on the original alignment of the neutrino mass states and whether the individual vacuum oscillation frequencies of those neutrinos are greater or less than $\omega_{\rm swap}$.  In the calculation presented here, which employs the normal neutrino mass hierarchy, a neutrino with vacuum oscillation frequency $\omega_{\rm V} > \omega_{\rm swap}$ will remain in the heavy, $\nu_3$, mass eigenstate.

Requiring that all neutrinos in the ensemble, across all emission trajectories, agree on $\omega_{\rm pr}$, as suggested by our results when the halo is included, admits an interesting solution.  In the single angle formulation, a neutrino which is precisely at the swap energy satisfies the criterion $\omega_{\rm pr} - \omega_{\rm V} = 0$.  In the region of the envelope where the swap is forming, e.g., where $\langle\vert \hat{H}_{\nu\nu}\vert\rangle \sim \omega_{\rm pr}$, there may be a significant difference between a neutrino's vacuum oscillation frequency and its instantaneous flavor oscillation frequency.  The latter arrises from dispersion in the forward scattering potentials along different emission trajectories.  This suggests a natural modification of the swap criterion which accommodates the additional dispersion present when the multi-dimensionality of the supernova environment is accounted for:
\begin{equation}
\omega_{\rm pr} -\left( \omega_{\rm V} + \Delta H_{\nu\nu} + \Delta H_{\rm e} + \Delta H_{\rm V} \right) = 0\ .
\label{newswap}
\end{equation}
This is directly equivalent to a shift in the swap frequency for each emission trajectory.  By grouping the dispersion terms with $\omega_{\rm pr}$, we define an effective, trajectory dependent swap frequency,
\begin{equation}
\tilde{\omega}_{\rm swap} = \omega_{\rm pr} - \left(  \Delta H_{\nu\nu} 
+ \Delta H_{\rm e} + \Delta H_{\rm V} \right)\ .
\label{wtilde}
\end{equation}

This explains neatly the effect seen in Figures 7 - 10.  In Figure~\ref{fig:MB7}, as the dispersion effect increases with increasing impact parameter, the mass state $3/2$ swap energy is pushed to much higher values, indeed crossing into the anti-neutrino sector, which corresponds to $\tilde{\omega}_{\rm swap} < 0$.  Figure~\ref{fig:SwapSurf} shows the isocontour which satisfies $\tilde\omega_{\rm swap} - \omega_{\rm V} = 0$ for the mass state $3/2$ swap.  This is plotted beneath the mass state occupation probability data.

When the isocontours for both swaps are considered, we recover the shape of the mass state 2 occupation probability which has produced the detectable swap feature for the $15\,\rm ms$ snapshot, shown in Figures~\ref{fig:MB15} and~\ref{fig:MB15B}.  In Figure~\ref{fig:SwapSurf2} both the isocontrous for the mass state $3/2$ and $2/1$ swaps are shown.  This is plotted alongside the mass state occupation probability data for mass state 2.  Because the mass state $2/1$ swap is formed by neutrino oscillations in the $\Delta m^2_{\odot}$ sector, the swap itself is formed when the neutrino self-coupling potential has a smaller magnitude than when the mass state $3/2$ swap is formed.  Consequently, the mass state $2/1$ swap is formed much further out in the envelope of the supernova.  At these larger radii, the dispersion effects which distort the swap surfaces are reduced, resulting a flavor swap that is less sensitive to the neutrino trajectory.  

%While the conserved constants of the motion for the collective neutrino oscillation are changed little by the addition of the neutrino halo, the spectral distortions created by the halo neutrinos can have major consequences for a detected signal.  Figures~\ref{fig:FBH} and~\ref{fig:FBNH} show the mass basis projection of the final neutrino states of neutrinos emitted $15\,\rm ms$ post core bounce for the halo and no-halo cases, respectively.  

\section{Conclusion}
We have made the first multi-angle calculation of neutrino flavor evolution in the supernova environment which includes the population of neutrinos scattered into the diffuse neutrino halo.  We have shown that there are qualitative differences in the neutrino oscillation patterns as a function of angle and energy.  These qualitative changes may also have potentially detectable consequences for a received neutronization burst signal.  

This calculation was made possible by a confluence of physical circumstances present during the neutronization burst of an O-Ne-Mg core collapse supernovae.  The combination of multi-angle suppression of neutrino flavor transformation deep within the envelope, and the precipitous drop in matter density just outside of the volume where multi-angle suppression ceases to operate, creates a unique situation where the halo neutrinos moving on inward directed trajectories are negligible at the radii where neutrino flavor transformation takes place.  This configuration allows for the inclusion of the neutrino halo within the present, initial value problem framework of computational models of supernova neutrino flavor transformation.

The results of our calculations show that the neutrino flavor swap, the clearest signature of neutrino collective oscillations, is a phenomenon dependent on the geometry of the neutrino flavor transformation environment, and hence of the geometry of the supernova envelope.  The geometric dependence of the swap energy is a feature of neutrino collective oscillations which has been completely overlooked by previous calculations.  The dependence of the halo neutrino flux on both the matter density and composition of the envelope now implies that the shape and spectral properties of the flavor swaps created in the explosion may also bear a dependence on the envelope as well.

Out results demonstrate the necessity of a self-consistent numerical approach in modeling collective oscillations in the fiercely nonlinear environment of stellar collapse.  For example, we found that while the small number of halo neutrinos has little effect on collective oscillations at $7\,\rm ms$, only $8\,\rm ms$ later these halo neutrinos significantly alter both the qualitative and quantitative process of flavor oscillation and swap formation.

The spectral distortions that the halo creates have potentially detectable consequences for a received neutronization neutrino burst signal here on Earth.  The halo shifts the apparent swap energies of the neutrino signal, sometimes across the boundary between $\nu$ and $\bar\nu$ sectors. This is accomplished without changing the overall spectral properties of the initial neutrino states.  Attempts~\cite{Cherry:2011bh,Cherry:2012lr} to reverse engineer a swap signal to extract information on the supernova environment must take account of the halo effect.

Finally this work gives a tantalizing glimpse of the new phenomenology which has emerged in spherical symmetry.  The Halo is, of course, a fundamentally multi-dimensional phenomena and these results strongly motivate attempts to expand the dimensionality of neutrino flavor transformation calculations.  Further, there are a multitude of different progenitor models which produce distinct signals~\cite{Cherry:2013yd} during the supernova explosion.  While we have presented an example of a single case here, a general solution for the effect of the scattered halo on flavor transformation in the explosion is intractable at present for later times and more massive supernova progenitors.  The distortion of the swap energy surface through emission angle space is a phenomenon that reveals how robust the collective oscillation modes of neutrinos can be.  Further, it exposes as false the fundamental assumption of the single-angle approximation: that individual neutrinos of the same energy and initial flavor state, following the same equation of motion, should be in identical flavor states.  If the distortion of spectral swaps persists in the presence of large numbers of inward directed neutrinos, it will be exacerbated by local sources of dispersion in the neutrino self-coupling potential, such as halo neutrino reflections off of turbulence driven cold matter accretion plumes.  This work raises the point that the coherent forward scattering of neutrinos and the Boltzmann transport of neutrinos do not belong in the separate camps to which they have apportioned in the supernova environment.  The inclusion of the neutrino halo in flavor transformation calculations is a zeroth order attempt to bridge this gap, and the results point directly to the need for full quantum kinetic treatment for the general case of neutrino transport in supernovae.

\section{Acknowledgments}
This work was supported in part by NSF grant PHY-09-70064 at UCSD and DOE Award DE-SC0008142 at UNM Albuquerque, and by the DOE Office of Science, the LDRD Program, Open Supercomputing at LANL, and the UC office of the President.  We would like to thank V. Cirigliano, H. Duan, Y.-Z. Qian, the Topical Collaboration for Neutrinos and Nucleosynthesis in Hot and Dense Matter at LANL, and the New Mexico Consortium.

\bibliography{./ONeMg_Halo_v2.bbl}

\begin{thebibliography}{22}
\expandafter\ifx\csname natexlab\endcsname\relax\def\natexlab#1{#1}\fi
\expandafter\ifx\csname bibnamefont\endcsname\relax
  \def\bibnamefont#1{#1}\fi
\expandafter\ifx\csname bibfnamefont\endcsname\relax
  \def\bibfnamefont#1{#1}\fi
\expandafter\ifx\csname citenamefont\endcsname\relax
  \def\citenamefont#1{#1}\fi
\expandafter\ifx\csname url\endcsname\relax
  \def\url#1{\texttt{#1}}\fi
\expandafter\ifx\csname urlprefix\endcsname\relax\def\urlprefix{URL }\fi
\providecommand{\bibinfo}[2]{#2}
\providecommand{\eprint}[2][]{\url{#2}}

\bibitem[{\citenamefont{{Cherry}
  et~al.}(2012{\natexlab{a}})\citenamefont{{Cherry}, {Carlson}, {Friedland},
  {Fuller}, and {Vlasenko}}}]{Cherry:2012uq}
\bibinfo{author}{\bibfnamefont{J.~F.} \bibnamefont{{Cherry}}},
  \bibinfo{author}{\bibfnamefont{J.}~\bibnamefont{{Carlson}}},
  \bibinfo{author}{\bibfnamefont{A.}~\bibnamefont{{Friedland}}},
  \bibinfo{author}{\bibfnamefont{G.~M.} \bibnamefont{{Fuller}}},
  \bibnamefont{and}
  \bibinfo{author}{\bibfnamefont{A.}~\bibnamefont{{Vlasenko}}},
  \bibinfo{journal}{Phys. Rev. Lett.} \textbf{\bibinfo{volume}{108}},
  \bibinfo{pages}{261104} (\bibinfo{year}{2012}{\natexlab{a}}),
  \eprint{1203.1607}.

\bibitem[{\citenamefont{{Sarikas} et~al.}(2012)\citenamefont{{Sarikas},
  {Tamborra}, {Raffelt}, {H{\"u}depohl}, and {Janka}}}]{Sarikas:2012qy}
\bibinfo{author}{\bibfnamefont{S.}~\bibnamefont{{Sarikas}}},
  \bibinfo{author}{\bibfnamefont{I.}~\bibnamefont{{Tamborra}}},
  \bibinfo{author}{\bibfnamefont{G.}~\bibnamefont{{Raffelt}}},
  \bibinfo{author}{\bibfnamefont{L.}~\bibnamefont{{H{\"u}depohl}}},
  \bibnamefont{and} \bibinfo{author}{\bibfnamefont{H.-T.}
  \bibnamefont{{Janka}}}, \bibinfo{journal}{ArXiv e-prints}
  (\bibinfo{year}{2012}), \eprint{1204.0971}.

\bibitem[{\citenamefont{{Freedman}}(1974)}]{Freedman:1974yq}
\bibinfo{author}{\bibfnamefont{D.~Z.} \bibnamefont{{Freedman}}},
  \bibinfo{journal}{Phys. Rev. D} \textbf{\bibinfo{volume}{9}},
  \bibinfo{pages}{1389} (\bibinfo{year}{1974}).

\bibitem[{\citenamefont{{Tubbs} and {Schramm}}(1975)}]{Tubbs:1975ve}
\bibinfo{author}{\bibfnamefont{D.~L.} \bibnamefont{{Tubbs}}} \bibnamefont{and}
  \bibinfo{author}{\bibfnamefont{D.~N.} \bibnamefont{{Schramm}}},
  \bibinfo{journal}{Astrophys. J.} \textbf{\bibinfo{volume}{201}},
  \bibinfo{pages}{467} (\bibinfo{year}{1975}).

\bibitem[{\citenamefont{{Bruenn} et~al.}(2010)\citenamefont{{Bruenn},
  {Mezzacappa}, {Hix}, {Blondin}, {Marronetti}, {Messer}, {Dirk}, and
  {Yoshida}}}]{Bruenn:2010qy}
\bibinfo{author}{\bibfnamefont{S.~W.} \bibnamefont{{Bruenn}}},
  \bibinfo{author}{\bibfnamefont{A.}~\bibnamefont{{Mezzacappa}}},
  \bibinfo{author}{\bibfnamefont{W.~R.} \bibnamefont{{Hix}}},
  \bibinfo{author}{\bibfnamefont{J.~M.} \bibnamefont{{Blondin}}},
  \bibinfo{author}{\bibfnamefont{P.}~\bibnamefont{{Marronetti}}},
  \bibinfo{author}{\bibfnamefont{O.~E.~B.} \bibnamefont{{Messer}}},
  \bibinfo{author}{\bibfnamefont{C.~J.} \bibnamefont{{Dirk}}},
  \bibnamefont{and}
  \bibinfo{author}{\bibfnamefont{S.}~\bibnamefont{{Yoshida}}},
  \bibinfo{journal}{ArXiv e-prints}  (\bibinfo{year}{2010}),
  \eprint{1002.4914}.

\bibitem[{\citenamefont{{M{\"u}ller} et~al.}(2010)\citenamefont{{M{\"u}ller},
  {Janka}, and {Dimmelmeier}}}]{Muller:2010kx}
\bibinfo{author}{\bibfnamefont{B.}~\bibnamefont{{M{\"u}ller}}},
  \bibinfo{author}{\bibfnamefont{H.}~\bibnamefont{{Janka}}}, \bibnamefont{and}
  \bibinfo{author}{\bibfnamefont{H.}~\bibnamefont{{Dimmelmeier}}},
  \bibinfo{journal}{The Astrophysical Journal Supplement}
  \textbf{\bibinfo{volume}{189}}, \bibinfo{pages}{104} (\bibinfo{year}{2010}),
  \eprint{1001.4841}.

\bibitem[{\citenamefont{{Brandt} et~al.}(2011)\citenamefont{{Brandt},
  {Burrows}, {Ott}, and {Livne}}}]{Brandt:2011lr}
\bibinfo{author}{\bibfnamefont{T.~D.} \bibnamefont{{Brandt}}},
  \bibinfo{author}{\bibfnamefont{A.}~\bibnamefont{{Burrows}}},
  \bibinfo{author}{\bibfnamefont{C.~D.} \bibnamefont{{Ott}}}, \bibnamefont{and}
  \bibinfo{author}{\bibfnamefont{E.}~\bibnamefont{{Livne}}},
  \bibinfo{journal}{Astrophys. J.} \textbf{\bibinfo{volume}{728}},
  \bibinfo{eid}{8} (\bibinfo{year}{2011}), \eprint{1009.4654}.

\bibitem[{\citenamefont{{M{\"u}ller} et~al.}(2012)\citenamefont{{M{\"u}ller},
  {Janka}, and {Heger}}}]{Muller:2012qy}
\bibinfo{author}{\bibfnamefont{B.}~\bibnamefont{{M{\"u}ller}}},
  \bibinfo{author}{\bibfnamefont{H.-T.} \bibnamefont{{Janka}}},
  \bibnamefont{and} \bibinfo{author}{\bibfnamefont{A.}~\bibnamefont{{Heger}}},
  \bibinfo{journal}{\apj} \textbf{\bibinfo{volume}{761}}, \bibinfo{eid}{72}
  (\bibinfo{year}{2012}), \eprint{1205.7078}.

\bibitem[{\citenamefont{{Bruenn} et~al.}(2013)\citenamefont{{Bruenn},
  {Mezzacappa}, {Hix}, {Lentz}, {Bronson Messer}, {Lingerfelt}, {Blondin},
  {Endeve}, {Marronetti}, and {Yakunin}}}]{Bruenn:2013fk}
\bibinfo{author}{\bibfnamefont{S.~W.} \bibnamefont{{Bruenn}}},
  \bibinfo{author}{\bibfnamefont{A.}~\bibnamefont{{Mezzacappa}}},
  \bibinfo{author}{\bibfnamefont{W.~R.} \bibnamefont{{Hix}}},
  \bibinfo{author}{\bibfnamefont{E.~J.} \bibnamefont{{Lentz}}},
  \bibinfo{author}{\bibfnamefont{O.~E.} \bibnamefont{{Bronson Messer}}},
  \bibinfo{author}{\bibfnamefont{E.~J.} \bibnamefont{{Lingerfelt}}},
  \bibinfo{author}{\bibfnamefont{J.~M.} \bibnamefont{{Blondin}}},
  \bibinfo{author}{\bibfnamefont{E.}~\bibnamefont{{Endeve}}},
  \bibinfo{author}{\bibfnamefont{P.}~\bibnamefont{{Marronetti}}},
  \bibnamefont{and} \bibinfo{author}{\bibfnamefont{K.~N.}
  \bibnamefont{{Yakunin}}}, \bibinfo{journal}{The Astrophysical Journal Letters}
  \textbf{\bibinfo{volume}{767}}, \bibinfo{eid}{L6} (\bibinfo{year}{2013}),
  \eprint{1212.1747}.

\bibitem[{\citenamefont{{Fischer} et~al.}(2010)\citenamefont{{Fischer},
  {Whitehouse}, {Mezzacappa}, {Thielemann}, and
  {Liebend{\"o}rfer}}}]{Fischer:2010lq}
\bibinfo{author}{\bibfnamefont{T.}~\bibnamefont{{Fischer}}},
  \bibinfo{author}{\bibfnamefont{S.~C.} \bibnamefont{{Whitehouse}}},
  \bibinfo{author}{\bibfnamefont{A.}~\bibnamefont{{Mezzacappa}}},
  \bibinfo{author}{\bibfnamefont{F.-K.} \bibnamefont{{Thielemann}}},
  \bibnamefont{and}
  \bibinfo{author}{\bibfnamefont{M.}~\bibnamefont{{Liebend{\"o}rfer}}},
  \bibinfo{journal}{Astronomy and Astrophysics} \textbf{\bibinfo{volume}{517}},
  \bibinfo{eid}{A80} (\bibinfo{year}{2010}), \eprint{0908.1871}.

\bibitem[{\citenamefont{{Wanajo} et~al.}(2011)\citenamefont{{Wanajo}, {Janka},
  and {M{\"u}ller}}}]{Wanajo:2011lr}
\bibinfo{author}{\bibfnamefont{S.}~\bibnamefont{{Wanajo}}},
  \bibinfo{author}{\bibfnamefont{H.-T.} \bibnamefont{{Janka}}},
  \bibnamefont{and}
  \bibinfo{author}{\bibfnamefont{B.}~\bibnamefont{{M{\"u}ller}}},
  \bibinfo{journal}{The Astrophysical Journal Letters}
  \textbf{\bibinfo{volume}{726}}, \bibinfo{eid}{L15} (\bibinfo{year}{2011}),
  \eprint{1009.1000}.

\bibitem[{\citenamefont{{Duan} et~al.}(2008{\natexlab{a}})\citenamefont{{Duan},
  {Fuller}, {Carlson}, and {Qian}}}]{Duan08}
\bibinfo{author}{\bibfnamefont{H.}~\bibnamefont{{Duan}}},
  \bibinfo{author}{\bibfnamefont{G.~M.} \bibnamefont{{Fuller}}},
  \bibinfo{author}{\bibfnamefont{J.}~\bibnamefont{{Carlson}}},
  \bibnamefont{and} \bibinfo{author}{\bibfnamefont{Y.}~\bibnamefont{{Qian}}},
  \bibinfo{journal}{Physical Review Letters} \textbf{\bibinfo{volume}{100}},
  \bibinfo{pages}{021101} (\bibinfo{year}{2008}{\natexlab{a}}),
  \eprint{0710.1271}.

\bibitem[{\citenamefont{{Lunardini} et~al.}(2008)\citenamefont{{Lunardini},
  {M{\"u}ller}, and {Janka}}}]{Lunardini08}
\bibinfo{author}{\bibfnamefont{C.}~\bibnamefont{{Lunardini}}},
  \bibinfo{author}{\bibfnamefont{B.}~\bibnamefont{{M{\"u}ller}}},
  \bibnamefont{and} \bibinfo{author}{\bibfnamefont{H.}~\bibnamefont{{Janka}}},
  \bibinfo{journal}{Phys. Rev. D} \textbf{\bibinfo{volume}{78}},
  \bibinfo{pages}{023016} (\bibinfo{year}{2008}), \eprint{0712.3000}.

\bibitem[{\citenamefont{{Cherry} et~al.}(2010)\citenamefont{{Cherry}, {Fuller},
  {Carlson}, {Duan}, and {Qian}}}]{Cherry:2010lr}
\bibinfo{author}{\bibfnamefont{J.~F.} \bibnamefont{{Cherry}}},
  \bibinfo{author}{\bibfnamefont{G.~M.} \bibnamefont{{Fuller}}},
  \bibinfo{author}{\bibfnamefont{J.}~\bibnamefont{{Carlson}}},
  \bibinfo{author}{\bibfnamefont{H.}~\bibnamefont{{Duan}}}, \bibnamefont{and}
  \bibinfo{author}{\bibfnamefont{Y.}~\bibnamefont{{Qian}}},
  \bibinfo{journal}{Phys. Rev. D} \textbf{\bibinfo{volume}{82}},
  \bibinfo{pages}{085025} (\bibinfo{year}{2010}), \eprint{1006.2175}.

\bibitem[{\citenamefont{{Nomoto}}(1984)}]{Nomoto84}
\bibinfo{author}{\bibfnamefont{K.}~\bibnamefont{{Nomoto}}},
  \bibinfo{journal}{Astrophys. J.} \textbf{\bibinfo{volume}{277}},
  \bibinfo{pages}{791} (\bibinfo{year}{1984}).

\bibitem[{\citenamefont{{Nomoto}}(1987)}]{Nomoto87}
\bibinfo{author}{\bibfnamefont{K.}~\bibnamefont{{Nomoto}}},
  \bibinfo{journal}{Astrophys. J.} \textbf{\bibinfo{volume}{322}},
  \bibinfo{pages}{206} (\bibinfo{year}{1987}).

\bibitem[{\citenamefont{{Scholberg}}(2011)}]{Scholberg:2011kj}
\bibinfo{author}{\bibfnamefont{K.}~\bibnamefont{{Scholberg}}}, in
  \emph{\bibinfo{booktitle}{APS Meeting Abstracts}} (\bibinfo{year}{2011}), p.
  \bibinfo{pages}{11006}.

\bibitem[{\citenamefont{{Scholberg}}(2012)}]{Scholberg:2012kj}
\bibinfo{author}{\bibfnamefont{K.}~\bibnamefont{{Scholberg}}},
  \bibinfo{journal}{Annual Review of Nuclear and Particle Science}
  \textbf{\bibinfo{volume}{62}}, \bibinfo{pages}{81} (\bibinfo{year}{2012}).

\bibitem[{\citenamefont{{Cherry}
  et~al.}(2012{\natexlab{b}})\citenamefont{{Cherry}, {Wu}, {Carlson}, {Duan},
  {Fuller}, and {Qian}}}]{Cherry:2012lr}
\bibinfo{author}{\bibfnamefont{J.~F.} \bibnamefont{{Cherry}}},
  \bibinfo{author}{\bibfnamefont{M.-R.} \bibnamefont{{Wu}}},
  \bibinfo{author}{\bibfnamefont{J.}~\bibnamefont{{Carlson}}},
  \bibinfo{author}{\bibfnamefont{H.}~\bibnamefont{{Duan}}},
  \bibinfo{author}{\bibfnamefont{G.~M.} \bibnamefont{{Fuller}}},
  \bibnamefont{and} \bibinfo{author}{\bibfnamefont{Y.-Z.}
  \bibnamefont{{Qian}}}, \bibinfo{journal}{\prd} \textbf{\bibinfo{volume}{85}},
  \bibinfo{eid}{125010} (\bibinfo{year}{2012}{\natexlab{b}}),
  \eprint{1109.5195}.

\bibitem[{\citenamefont{{Duan} et~al.}(2008{\natexlab{b}})\citenamefont{{Duan},
  {Fuller}, and {Qian}}}]{Duan:2008qy}
\bibinfo{author}{\bibfnamefont{H.}~\bibnamefont{{Duan}}},
  \bibinfo{author}{\bibfnamefont{G.~M.} \bibnamefont{{Fuller}}},
  \bibnamefont{and} \bibinfo{author}{\bibfnamefont{Y.}~\bibnamefont{{Qian}}},
  \bibinfo{journal}{Phys. Rev. D} \textbf{\bibinfo{volume}{77}},
  \bibinfo{pages}{085016} (\bibinfo{year}{2008}{\natexlab{b}}),
  \eprint{0801.1363}.

\bibitem[{\citenamefont{{Cherry} et~al.}(2011)\citenamefont{{Cherry}, {Wu},
  {Carlson}, {Duan}, {Fuller}, and {Qian}}}]{Cherry:2011bh}
\bibinfo{author}{\bibfnamefont{J.~F.} \bibnamefont{{Cherry}}},
  \bibinfo{author}{\bibfnamefont{M.-R.} \bibnamefont{{Wu}}},
  \bibinfo{author}{\bibfnamefont{J.}~\bibnamefont{{Carlson}}},
  \bibinfo{author}{\bibfnamefont{H.}~\bibnamefont{{Duan}}},
  \bibinfo{author}{\bibfnamefont{G.~M.} \bibnamefont{{Fuller}}},
  \bibnamefont{and} \bibinfo{author}{\bibfnamefont{Y.-Z.}
  \bibnamefont{{Qian}}}, \bibinfo{journal}{Phys. Rev. D}
  \textbf{\bibinfo{volume}{84}}, \bibinfo{eid}{105034} (\bibinfo{year}{2011}),
  \eprint{1108.4064}.

\bibitem[{\citenamefont{{Cherry} and {Duan}}(2013)}]{Cherry:2013yd}
\bibinfo{author}{\bibfnamefont{J.~F.} \bibnamefont{{Cherry}}} \bibnamefont{and}
  \bibinfo{author}{\bibfnamefont{H.}~\bibnamefont{{Duan}}}
  (\bibinfo{year}{2013}), \bibinfo{note}{in preparation}.

\end{thebibliography}
%\bibliography{/Users/JJ/Documents/Bibliography/allref}
%\bibliography{./ONeMg_Halo_v2.bbl}

\end{document}